\documentclass[journal,11pt,a4paper]{IEEEtran}
\DeclareMathSizes{11}{9.8}{6}{4}
\usepackage[english]{babel}
\usepackage{flushend}
\usepackage{amssymb}
\usepackage[cmex10]{amsmath}
\usepackage{overpic}
\usepackage{algorithm}
\usepackage{algorithmic}
\usepackage{multirow}
\usepackage{array}
\usepackage{graphicx}
\usepackage{bm}
\usepackage[caption=false]{subfig}
\usepackage{color}
\usepackage{tikz}
\usepackage{epstopdf}
\usepackage{stfloats}
\usepackage{cite}
\usepackage{tikz}
\usepackage{amsmath}
\usepackage{psfrag}
\usepackage{acronym}
\usepackage{enumerate}
\usetikzlibrary{arrows}
\usetikzlibrary{decorations}

\definecolor{BLUE}{rgb}{0,0,1}

\newcommand{\tr}[1]{{\rm tr}\left\{#1\right\}}

\acrodef{aoa}[AOA]{angle-of-arrival}
\acrodef{bcrb}[BCRB]{Bayesian Cram\'{e}r-Rao bound}
\acrodef{bfim}[BFIM]{Bayesian Fisher information matrix}
\acrodef{bp}[BP]{belief propagation}
\acrodef{bs}[BS]{base station}
\acrodef{cdi}[CDI]{cooperative dilution intensity}
\acrodef{cir}[CIR]{channel impulse response}
\acrodef{cl}[CL]{cooperative localization}
\acrodef{cp}[CP]{cyclic prefix}
\acrodef{crb}[CRB]{Cram\'{e}r-Rao bound}
\acrodef{crlb}[CRLB]{Cram\'{e}r-Rao lower bound}
\acrodef{dft}[DFT]{discrete Fourier transform}
\acrodef{dof}[DoF]{degree of freedom}
\acrodef{dpeb}[DPEB]{directional position error bound}
\acrodef{fim}[FIM]{Fisher information matrix}
\acrodef{efim}[EFIM]{equivalent Fisher information matrix}
\acrodef{hogs}[HOGS]{hybrid orthogonal-Gaussian signalling}
\acrodef{hsug}[HSUG]{hybrid semi-unitary--Gaussian}
\acrodef{ici}[ICI]{information coupling intensity}
\acrodef{icrb}[ICRB]{inverse CRB}
\acrodef{iid}[i.i.d.]{independently and identically distributed}
\acrodef{im}[IM]{index modulation}
\acrodef{isac}[ISAC]{integrated sensing and communications}
\acrodef{los}[LoS]{line-of-sight}
\acrodef{mse}[MSE]{mean-squared error}
\acrodef{ofdm}[OFDM]{orthogonal frequency-division multiplexing}
\acrodef{pdf}[PDF]{probability density function}
\acrodef{peb}[PEB]{position error bound}
\acrodef{speb}[SPEB]{squared position error bound}
\acrodef{pll}[PLL]{phase-locked loop}
\acrodef{psk}[PSK]{phase shift keying}
\acrodef{p2p}[P2P]{point-to-point}
\acrodef{qam}[QAM]{quadrature amplitude modulation}
\acrodef{rbs}[RBS]{reference broadcast synchronization}
\acrodef{rhs}[RHS]{right hand side}
\acrodef{rii}[RII]{ranging information intensity}
\acrodef{rss}[RSS]{received signal strength}
\acrodef{rc}[RC]{ranging coefficient}
\acrodef{speb}[SPEB]{squared position error bound}
\acrodef{toa}[TOA]{time-of-arrival}
\acrodef{tdoa}[TDOA]{time-difference-of-arrival}
\acrodef{tpsn}[TPSN]{time synchronization protocol for sensor network}
\acrodef{vmp}[VMP]{variational message passing}
\acrodef{wsn}[WSN]{wireless sensor network}
\acrodef{efim}[EFIM]{equivalent Fisher information matrix}
\acrodef{dio}[DIO]{distance-information-only}
\acrodef{aio}[AIO]{angle-information-only}
\acrodef{saaf}[SAAF]{squared array aperture function}
\acrodef{snc}[S\&C]{sensing and communications}
\acrodef{uoa}[UOA]{uniformly oriented array}
\acrodef{rgg}[RGG]{random geometric graph}
\acrodef{snr}[SNR]{signal-to-noise ratio}
\acrodef{eoc}[EoC]{efficiency of cooperation}
\acrodef{npi}[NPI]{nominal position information}
\acrodef{gnss}[GNSS]{global navigation satellite system}
\acrodef{mimo}[MIMO]{multiple-input multiple-output}
\acrodef{siso}[SISO]{single-input single-output}
\acrodef{mcs}[MCS]{minimally constrained system}
\acrodef{zzb}[ZZB]{Ziv-Zakai bound}
\acrodef{wwb}[WWB]{Weiss-Weinstein lower bound}
\acrodef{nlos}[NLOS]{non-light-of-sight}
\acrodef{mmse}[MMSE]{minimum mean squared error}
\acrodef{uav}[UAV]{unmanned aerial vehicle}
\acrodef{ppp}[PPP]{Poisson point process}
\acrodef{bpp}[BPP]{binomial point process}
\acrodef{cln}[CLN]{cooperative location-aware network}
\acrodef{pdr}[PDR]{pedestrian dead reckoning}
\acrodef{ml}[ML]{maximum likelihood}
\acrodef{map}[MAP]{maximum \textit{a posteriori}}
\acrodef{kkt}[KKT]{Karush-Kuhn-Tucker}
\acrodef{st}[ST]{subspace tradeoff}
\acrodef{drt}[DRT]{deterministic-random tradeoff}
\acrodef{ustm}[USTM]{unitary space-time modulation}
\acrodef{kld}[KLD]{Kullback-Leibler divergence}
\acrodef{cfar}[CFAR]{constant false-alarm rate}
\acrodef{sdmcdf}[SDMC-DF]{state-dependent memoryless channel with delayed feedback}
\acrodef{ba}[B-A]{Blahut-Arimoto}
\acrodef{bpsk}[BPSK]{binary phase-shift keying}
\acrodef{tx}[Tx]{transmitter}
\acrodef{rx}[Rx]{receiver}
\acrodef{itu}[ITU]{international telecommunication union}
\acrodef{rms}[RMS]{root-mean-square}
\acrodef{gnss}[GNSS]{global navigation satellite system}
\acrodef{acf}[ACF]{autocorrelation function}
\acrodef{psd}[PSD]{power spectral density}
\acrodef{awgn}[AWGN]{additive white Gaussian noise}
\acrodef{mac}[MAC]{multiple access channel}

\acrodefplural{dof}[DoFs]{degrees of freedom}
\acrodefplural{drt}[DRTs]{deterministic-random tradeoffs}
\acrodefplural{psd}[PSDs]{power spectral densities}

\usepackage{winsnotation}

\title{From Torch to Projector: Fundamental Tradeoff of Integrated Sensing and Communications}
\author{Yifeng Xiong, Fan Liu, Kai Wan, Weijie Yuan, Yuanhao Cui, and Giuseppe Caire}

\begin{document}
\maketitle

\begin{abstract}
\Ac{snc} have been historically developed in parallel. In recent decade, they have been evolving from separation to integration, giving rise to the \ac{isac} paradigm, that has been recognized as one of the six key 6G usage scenarios. Despite the plethora of research works dedicated to \ac{isac} signal processing, the fundamental performance limits of \ac{snc} remain widely unexplored in an \ac{isac} system. In this tutorial paper, we attempt to summarize the recent research findings in characterizing the performance boundary of \ac{isac} systems and the resulting \ac{snc} tradeoff from an information-theoretical viewpoint. We begin with a folklore ``torch metaphor" that depicts the resource competition mechanism of \ac{snc}. Then, we elaborate on the fundamental capacity-distortion (C-D) theory, indicating the incompleteness of this metaphor. Towards that end, we further elaborate on the \ac{snc} tradeoff by discussing a special case within the C-D framework, namely the Cram\'er-Rao bound (CRB)-rate region. In particular, \ac{snc} have preference discrepancies over both the subspace occupied by the transmitted signal and the adopted codebook, leading to a ``projector metaphor" complementary to the \ac{isac} torch analogy. We also present two practical design examples by leveraging the lessons learned from fundamental theories. Finally, we conclude the paper by identifying a number of open challenges.
\end{abstract}

\section{Introduction}
\IEEEPARstart{I}{SAC} has recently been recognized by the \ac{itu} as one of the vertices supporting the 6G hexagon of usage scenarios \cite{ITU2023}. By relying on unified hardware platforms and radio waveforms, such integration enables resource-efficient cooperation between \ac{snc} subsystems, supporting various emerging applications including vehicular-to-everything (V2X) networking, extended reality (XR), and digital twins. 

Early endeavours on \ac{isac} system design aim to extend the capability of existing infrastructures, hence developed into \emph{sensing-centric} and \emph{communication-centric} paradigms. Representative sensing-centric schemes include communicating using radar sidelobes and the permutation \ac{dof} in the waveform-antenna mapping of \ac{mimo} radars \cite{8828023}. Communication-centric schemes are exemplified by sensing relying on \ac{ofdm} waveforms \cite{sturm2011waveform}. These schemes do not target the optimal \ac{snc} performance. To push the benefit of \ac{isac} (termed as the ``integration gain'' \cite{isac_network}) to its limit, \emph{joint design} schemes emerged more recently \cite{crb_radcom}, which aim at conceiving novel joint signaling strategies from the ground-up, capable of accomplishing both tasks simultaneously. A natural, but easily overlooked question that arises is: What is the fundamental limit of the integration gain?

If we focus back on conventional individual \ac{snc} systems, we see that their fundamental performance limits originate from the resource budget. For example, the celebrated Shannon capacity formula for the scalar Gaussian channel exactly expresses the dependency of the communication performance on the available power and bandwidth. However, the performance limit of \ac{isac} systems is vastly different. In its essence, the problem of \ac{isac} system design is a multi-objective optimization problem. To elaborate, separated \ac{snc} design can be viewed as a special case of \ac{isac} design, namely employing a time-sharing strategy. By utilizing the synergies between \ac{snc}, more favorable performance can be achieved. However, besides some rare occasions, it is highly unlikely that \ac{snc} would achieve their optimal performance at the same time, suggesting the existence of a fundamental \ac{snc} tradeoff in an ISAC system. Such a tradeoff may be framed by the Pareto boundary of the multi-objective \ac{isac} system design problem.

For decades, \ac{snc} have been regarded as an information-theoretical ``odd couple" that are mutually intertwined in profound ways \cite{odd_couple}. At large, the \ac{snc} tradeoff may be understood from the perspective of a {\it{signal-and-system}} duality. For instance, let us consider a simple linear Gaussian model given by
\begin{equation}
\RM{Y}=\RM{H}\RM{X}+\RM{Z}
\end{equation}
with $\RM{X}$, $\RM{Y}$, $\RM{H}$ and $\RM{Z}$ being the transmitted \ac{isac} signal, the received signal, the channel, and the noise, respectively. For communication, the task is to decode the message encoded in $\RM{X}$, whereas for sensing, the task is to extract information from the channel $\RM{H}$. In light of this, one may view $\RM{H}$ as the ``transmitted signal'' from the environment to be sensed, while viewing $\RM{X}$ as the ``channel' that this ``signal'' would pass through. 

While the connection between estimation and information theories has been well-studied in the context of, e.g., I-MMSE equation \cite{immse}, the fundamental tradeoff still remains widely open in general. To this end, this article will summarize current understandings of different components in the \ac{snc} tradeoff in a coherent manner, through the prism of information theory \cite{cd, cdmac, CD_TIT,10147248,CE_detection}. We shall commence with a folklore ``torch metaphor'' depicting the resource competition between \ac{snc} subsystems, followed by the general capacity-distortion theory suggesting the incompleteness of this metaphor. We then introduce the CRB-rate region which clearly indicates that the \ac{snc} tradeoff is two-fold: Apart from resources, \ac{snc} subsystems would also have different preferences on the input distribution. For each component in this tradeoff, we provide design examples illustrating its practical implications. Finally, we conclude this paper with some open challenges.

\section{The Torch Metaphor}
\begin{figure}[t]
    \centering
    \includegraphics[width=.45\textwidth]{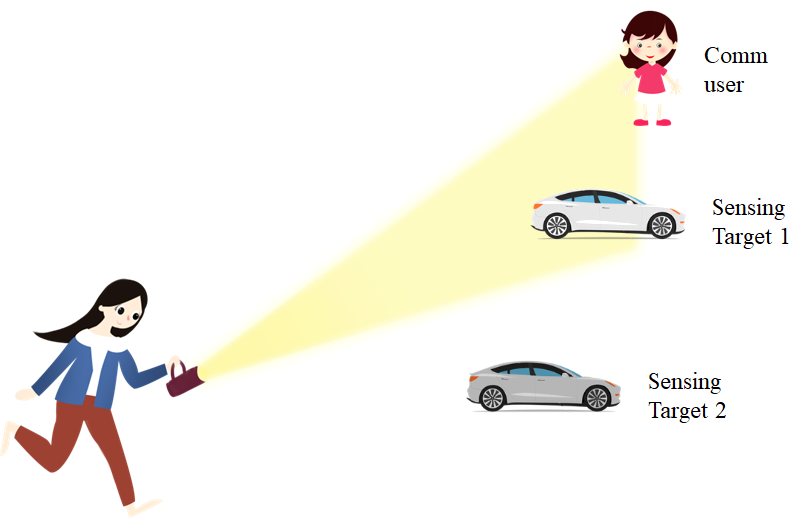}
    \caption{Graphical illustration of the torch metaphor.}
    \label{fig:torch_metaphor_illustration}
    \vspace{-3mm}
\end{figure}

The fundamental \ac{snc} tradeoff in \ac{isac} has been exemplified (implicitly or explicitly) by the ``torch metaphor'' in some works \cite{10217169, crb_radcom}, as illustrated in Fig.~\ref{fig:torch_metaphor_illustration}. In this picture, the role played by the \ac{isac} \ac{bs} resembles a child holding a torch: If she points the torch towards the communication user, the user would receive the message, while the sensing target is left in the dark and hence can hardly be seen. On the other hand, if the target is maximally illuminated, the user would receive very weak signals, resulting in highly noise-corrupted messages. This metaphor is intuitive, and it provides us with the following basic understandings of the \ac{snc} tradeoff:
\begin{itemize}
    \item Power allocation across orthogonal or quasi-orthogonal dimensions (in the case of Fig.~\ref{fig:torch_metaphor_illustration} it is space, or angle, but it could be time, or frequency) offers an immediate way to tradeoff communication and sensing performance;
    \item Since we are using a unified signal, when the communication user and the sensing target are ``close to each other'', the \ac{snc} tradeoff becomes less prominent.
\end{itemize}

Inspired by these intuitions, most existing research contributions on \ac{isac} system design focus on power allocation in a generalized sense, including beamforming in \ac{mimo} systems \cite{crb_radcom} and subcarrier power allocation in \ac{ofdm} systems \cite{8561147}. Despite the effectiveness of these techniques, we would still wonder, whether the torch metaphor depicts the full picture of the \ac{snc} tradeoff. 

Recently, it has been found that the torch metaphor does cover the full picture, when the sensing task is to detect the presence of a potential target, under some specific choice of sensing performance metrics \cite{CE_detection}. Specifically, in this scenario, during the $n$-th channel use, the \ac{isac} \ac{bs} would determine the absence/presence of the target (represented by a state $\rv{\eta}\in\{0,1\}$) based on an echo $\RM{Y}_{{\rm s},n}$, while the user aims for decoding the information conveyed in the transmitted \ac{isac} signal $\RM{X}_n$, based on its received signal $\RM{Y}_{{\rm c},n}$. Both the echo and the user-received signal are contaminated by circularly symmetric Gaussian noises, denoted by $\RM{Z}_{{\rm s},n}$ and $\RM{Z}_{{\rm c},n}$, respectively. Such a scenario may be characterized as
\begin{subequations}\label{model_detection}
\begin{align}
\RM{Y}_{{\rm c},n} &= \M{H}_{\rm c}\RM{X}_n+\RM{Z}_{{\rm c},n}, \\ 
\RM{Y}_{{\rm s},n} &= \rv{\eta}\M{H}_{\rm s}\RM{X}_n+\RM{Z}_{{\rm s},n},
\end{align}
\end{subequations}\noindent
where $n=1,2,\dotsc,N$, $N$ is the coding block length, $\M{H}_{\rm c}$ denotes the communication channel, while $\M{H}_{\rm s}$ represents the target response matrix (also referred to as the sensing channel \cite{CD_TIT,10147248,CE_detection}). For communication, a natural performance metric is the communication rate given by
$$
R=\lim_{N\rightarrow \infty} \frac{1}{N}\log M_N,
$$  
under the assumption that the decoding error probability vanishes as $N$ tends to infinity, and $M_N$ denotes the size of the communication codebook. For sensing, the performance metric chosen in \cite{CE_detection,CE_detection_TIT} is the error exponent\footnote{The error exponent is related to a class of more frequently used sensing performance metrics, i.e. statistical divergences, via the large deviation theory \cite{cover}.} defined as
\begin{equation}
E = \lim_{N\rightarrow \infty} \frac{1}{N}\log\frac{1}{\delta_N},
\end{equation}\noindent
where $\delta_N$ denotes the maximum detection error probability over all codewords and target states, namely
$$
\delta_N = \max_{w\in[M_N]}\max_{s\in\{0,1\}} \mathbb{P}\left\{\hat{S}\left(\RM{Y}_{\rm s}^N,\RM{X}^N(\rv{W})\right)\neq \rv{\eta}|\rv{\eta}=s,\rv{W}=w\right\},
$$
with $\rv{W}\in[M_N]$ being the encoded message, $\RM{Y}_{\rm s}^N\!=\!\{\RM{Y}_{{\rm s},1},\RM{Y}_{{\rm s},2},\dotsc,\RM{Y}_{{\rm s},N}\}$, and $\RM{X}^N(W)\!=\!\{\RM{X}_1,\RM{X}_2,\dotsc,\RM{X}_N\}$. Under the aforementioned setting, the set constituted by all achievable $(R,E)$ pairs (termed as the rate-exponent region) is shown to be characterized by \cite{CE_detection}
\begin{subequations}\label{re_region}
\begin{align}
R &\leq \log \left|\M{I}+\sigma_{\rm c}^{-2}\M{H}_{\rm c}\widetilde{\M{R}}_{\RM{X}}\M{H}_{\rm c}^{\rm H}\right|,\\ 
E &\leq \frac{1}{4}{\rm Tr}\left\{\sigma_{\rm s}^{-2}\M{H}_{\rm s}\widetilde{\M{R}}_{\RM{X}}\M{H}_{\rm s}^{\rm H}\right\},
\end{align}
\end{subequations}\noindent
where $\widetilde{\M{R}}_{\RM{X}}=\mathbb{E}\left\{\frac{1}{N}\sum_{n=1}^{N}\RM{X}_n\RM{X}_n^{\rm H}\right\}$ is the statistical covariance matrix of the transmitted \ac{isac} signal, satisfying a power budget constraint ${\rm Tr}\{\widetilde{\M{R}}_{\RM{X}}\}\leq P$. 

From \eqref{re_region} we observe a concrete version of the torch metaphor. Apparently, the communication-optimal $\widetilde{\M{R}}_{\RM{X}}$ would have its eigenvectors matching those of $\M{H}_{\rm c}$, while the eigenvalues determined by the water-filling power allocation \cite{cover}. By contrast, the sensing-optimal $\widetilde{\M{R}}_{\RM{X}}$ would be aligned with the dominant eigenvector (possibly having multiplicity larger than $1$) of $\M{H}_{\rm s}$ \cite{CE_detection}. Therefore, as long as the eigenspaces of $\M{H}_{\rm s}$ and $\M{H}_{\rm c}$ do not match, or the water-filling strategy does not concentrate the power on the dominant eigenvector, there would be a \ac{snc} tradeoff. Furthermore, \eqref{re_region} also gives us an explicit sense in which the communication user and the sensing target are ``close to each other'': their distance may be characterized by a measure of discrepancy between the ``communication subspace'' ${\rm span}(\M{H}_{\rm c})$ and the ``sensing subspace'' -- the dominant eigenvector of $\M{H}_{\rm s}$. 
In light of this, such a tradeoff is termed as the \ac{st} in \cite{CD_TIT}, and thus the torch metaphor may be represented concisely as the statement of 
\begin{equation}\label{torch_formulation}
\mathrm{S\& C}~{\rm tradeoff} = \mathrm{ST}.
\end{equation}

To further reveal the nature of the \ac{st}, let us consider a tangible example, in which the \ac{bs} is equipped with collocated transmit and receive uniform linear antenna arrays (half-wavelength spacing) of the size $N_{\rm s}=M=10$, the user has a single antenna (i.e. $N_{\rm c}=1$), while the sensing target is point-like. In this scenario, the communication and sensing channels are given by
$$
\M{H}_{\rm c} = \alpha_{\rm c} \V{a}^{\rm H}(\theta_{\rm c}),~~\M{H}_{\rm s} = \alpha_{\rm s} \V{a}(\theta_{\rm s})\V{a}^{\rm H}(\theta_{\rm s}),
$$
where $\alpha_{\rm c}$ and $\alpha_{\rm s}$ denote the scalar channel coefficients, $\theta_{\rm c}$ and $\theta_{\rm s}$ denote the bearing angle of the user and the target relative to the \ac{bs}, respectively, while $\V{a}(\theta)$ is the array steering vector given by
$$
\V{a}(\theta) = [1,~e^{j\pi\sin(\theta)},~e^{j2\pi\sin(\theta)},\dotsc,e^{j\pi(N_{\rm s}-1)\sin(\theta)}]^{\rm T}.
$$
Correspondingly, the characterization of the rate-exponent region \eqref{re_region} can then be simplified as
\begin{subequations}\label{re_region_example}
\begin{align}
\V{r}(\lambda) &\!=\! \mathop{\rm argmax}_{\V{v}:\|\V{v}\|=1}~(1\!-\!\lambda)|\V{v}^{\rm H}\V{a}(\theta_{\rm c})|^2\!+\!\lambda|\V{v}^{\rm H}\V{a}(\theta_{\rm s})|^2,~\lambda\!\in\![0,1]\\
R(\lambda)&\!\leq\! \log (1+P\sigma_{\rm c}^{-2}|\alpha_{\rm c}|^2|\V{r}^{\rm H}(\lambda)\V{a}(\theta_{\rm c})|^2),\\
E(\lambda)&\!\leq\! \frac{1}{4}{\rm Tr}\left\{P\sigma_{\rm s}^{-2} |\alpha_{\rm s}|^2|\V{r}^{\rm H}(\lambda)\V{a}(\theta_{\rm s})|^2 \right\},
\end{align}
\end{subequations}
where $\lambda$ is a parameter controlling the \ac{snc} tradeoff. The communication and sensing \acp{snr} may be expressed as
$$
{\rm SNR}_{\rm c} = P\|\M{H}_{\rm c}\|_{\rm F}^2\sigma_{\rm c}^{-2},~~{\rm SNR}_{\rm s} = P\|\M{H}_{\rm s}\|_{\rm F}^2\sigma_{\rm s}^{-2}.
$$

\begin{figure}
\centering
\includegraphics[width=.45\textwidth]{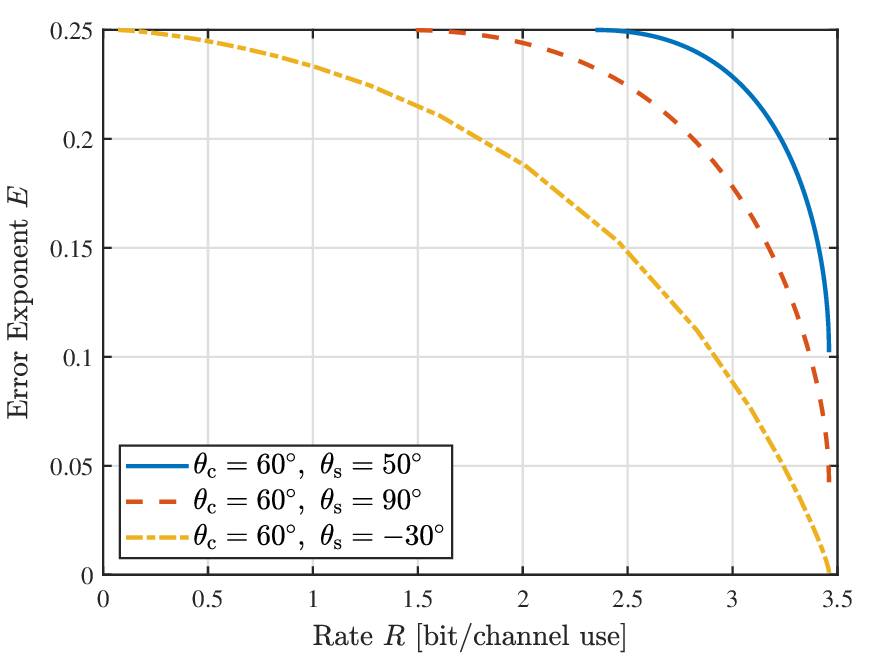}
\caption{The rate-exponent region of the single-antenna user, point-like target scenario, with communication and sensing \acp{snr} equal to $10$dB and $0$dB, respectively.}
\label{fig:re_region}
\vspace{-3mm}
\end{figure}

\begin{figure}
\centering
\subfloat[$\theta_{\rm c}=60^{\circ},~\theta_{\rm s}=90^{\circ}$]{
\centering
\includegraphics[width=.45\textwidth]{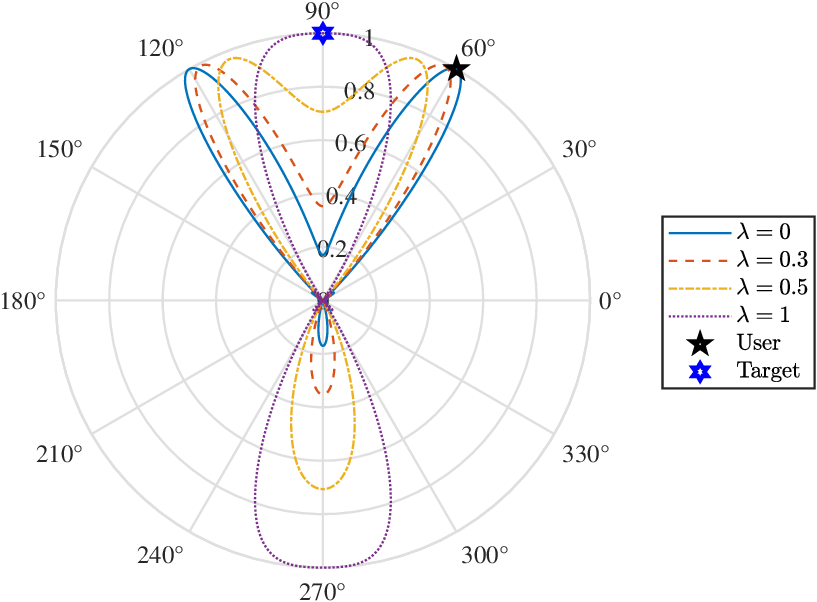}
\label{fig:polar_re_90}
} \\
\subfloat[$\theta_{\rm c}=60^{\circ},~\theta_{\rm s}=-30^{\circ}$]{
\centering
\includegraphics[width=.45\textwidth]{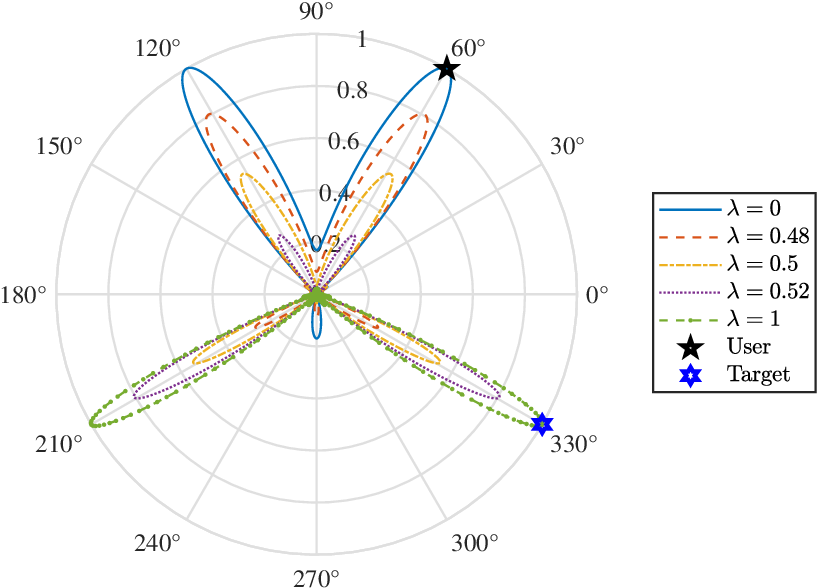}
\label{fig:polar_re_210}
}
\caption{Beamspace illustration of the subspace tradeoff (corresponding to the rate-exponent regions in Fig.~\ref{fig:re_region}). The ``\ac{isac} torch'' can simultaneously illuminate both the user and the target in case (a), while it has to apply a power-splitting strategy in case (b).}
\label{fig:polars}
\vspace{-3mm}
\end{figure}

Using \eqref{re_region_example}, we may readily obtain the rate-exponent regions for given configurations of $(\theta_{\rm c},\theta_{\rm s})$, as portrayed in Fig.~\ref{fig:re_region}. Note that the closeness between the \ac{snc} subspaces in this scenario is characterized by the intersection angle between $\V{a}(\theta_{\rm c})$ and $\V{a}(\theta_{\rm s})$. We may observe from Fig.~\ref{fig:re_region} that the \ac{snc} tradeoff becomes more prominent as the angular separation between the target and the user increases. More intuitively, as can be seen from Fig.~\ref{fig:re_region}, in the $\theta_{\rm c}=60^{\circ}$, $\theta_{\rm s}=90^{\circ}$ scenario, the sensing-optimal and communication-optimal beam patterns (corresponding to $\lambda=0$ and $\lambda=1$, respectively) have a large overlap. By contrast, in the $\theta_{\rm c}=60^{\circ}$, $\theta_{\rm s}=-30^{\circ}$ scenario, the sensing- and communication-optimal beam patterns are almost orthogonal to each other. This corroborates our intuition about the \ac{st}, that more synergies between \ac{snc} tasks would be witnessed as the corresponding subspaces become closer to each other. In the language of the torch metaphor, we may say that the ``\ac{isac} torch'' in the case of Fig.~\ref{fig:polar_re_90} can simultaneously illuminate both the user and the target, while it has to apply a power-splitting strategy in the case of Fig.~\ref{fig:polar_re_210}.

Now that we see \eqref{torch_formulation} holds true for the target presence detection problem under the sensing metric of error exponent, we would naturally ask in the sequel that:
\begin{itemize}
\item Does \eqref{torch_formulation} hold in general?
\item If not, what is the key condition for \eqref{torch_formulation} to hold? Is it the metric chosen (error exponent), the specific type of the sensing task (target detection), or both?
\end{itemize}

\section{Capacity-Distortion Theory}
To answer these questions, one has to rely on a more general analytical framework, that applies to both estimation and detection tasks and can handle favorably all reasonable sensing performance metrics. The capacity-distortion theory \cite{cd, cdmac, CD_TIT} is conceived in the hope of building such a universal framework.

\subsection{Rate-Distortion Region}
As the name suggests, the capacity-distortion theory investigates the tradeoff between communication capacity and sensing distortion. Originally proposed by Shannon in the context of rate-distortion theory for lossy data compression \cite{5311476}, distortion refers to a wide range of functions taking the form of $d(\RV{\eta},\hat{\RV{\eta}})$, whose inputs are the true value of some quantity (in sensing problems, the sensing parameter) $\RV{\eta}$ and its estimate $\hat{\RV{\eta}}$. Due to the randomness of the communication message, the \ac{isac} waveform (codeword) $\RM{X}^N$ is also random. To reflect the sensing performance over a relatively long period of time, a common practice is to use the expectation of the distortion, instead of its instantaneous values, as the performance metric. For example, in estimation tasks, the squared Euclidean distance $d(\RV{\eta},\hat{\RV{\eta}})=\|\RV{\eta}-\hat{\RV{\eta}}\|^2$ is a widely applied distortion function, whose expectation is the \ac{mse}. In binary (e.g. target presence) detection problems in which the task is to determine the value of a binary variable $\rv{\eta}\in\{0,1\}$, a valid distortion function is the Hamming distance given by $d(\rv{\eta},\hat{\rv{\eta}}) = \rv{\eta}\oplus \hat{\rv{\eta}}$, whose expectation is related to commonly used detection metrics, including the detection probability $P_{\rm D}$ and the false alarm rate $P_{\rm FA}$, as follows:
\begin{align}
&\mathbb{E}\{\rv{\eta}\oplus\hat{\rv{\eta}}\} \nonumber \\
&\hspace{3mm}=(1\oplus 1)\mathbb{P}\{\hat{\rv{\eta}}=1|\rv{\eta}=1\}+(0\oplus 0)\mathbb{P}\{\hat{\rv{\eta}}=0|\rv{\eta}=0\}\nonumber\\
&\hspace{7mm}+(1\oplus 0)\mathbb{P}\{\hat{\rv{\eta}}=1|\rv{\eta}=0\}+(0\oplus 1)\mathbb{P}\{\hat{\rv{\eta}}=0|\rv{\eta}=1\}\nonumber\\
&\hspace{3mm}=1-P_{\rm D}+P_{\rm FA}.
\end{align}   
Note that for \ac{cfar} detectors based on the Neyman-Pearson criterion with fixed $P_{\rm FA}$ \cite{kay1998fundamentals3}, minimizing the Hamming distance is equivalent to maximizing the detection probability $P_{\rm D}$.

Besides capacity and distortion, there is yet another important ingredient in the capacity-distortion theory, namely the transmission cost. To elaborate, not all \ac{isac} waveforms (or codewords) cost equally in terms of wireless resources. For example, the points in \ac{qam} constellations having different amplitudes would yield different power consumptions. Apparently, as the overall resource budget increases, \ac{snc} performances can be simultaneously enhanced. Therefore, one has to discuss the capacity-distortion tradeoff under a specific resource budget. Similar to the sensing distortion, in order to take into account the randomness of the codewords, we typically use the expectation of the resource cost over all possible codewords as the measure of transmission cost.

Once the resource budget is given and the sensing distortion metric is chosen, the \ac{snc} tradeoff is expressed in terms of the largest achievable rate-distortion region. Formally speaking, given an expected resource budget $B$, the rate-distortion-cost triple $(R,D,B)$ is said to be achievable (in the infinite block length regime), if there exists a sequence of $(2^{NR},R)$ codes $\{\RM{X}^N|N\in\mathbb{N}\}$ encoding the message $\rv{W}\in\{1,\dotsc,2^{NR}\}$, and a state estimator function $\hat{\RV{\eta}}: (\RM{Y}_{\rm s},\RM{X}^N) \mapsto \hat{\RV{\eta}}$, such that the following holds \cite{CD_TIT}
\begin{subequations}\label{capacity_distortion_cost}
\begin{align}
&\mathbb{E}\big\{d\big(\RV{\eta}^N,\hat{\RV{\eta}}^N\big)\big\}\leq D,\\
&\mathbb{E}\{b(\RM{X}^N)\}\leq B,\\
&P_{\rm e}^{(N)}:=\frac{1}{2^{NR}}\sum_{i=1}^{2^{NR}}\mathbb{P}\{\hat{\rv{W}}\neq i|\rv{W}=i\}\rightarrow 0,
\end{align}
\end{subequations}
as $N\rightarrow \infty$, where $b(\cdot)$ denotes the instantaneous cost of single codeword. In addition, for generality, the values of the sensing parameters are allowed to vary across time, denoted by $\RV{\eta}_i$ at the $i$-th channel use, constituting a parameter sequence $\RV{\eta}^N$. The capacity-distortion function given a specific resource budget $B$ is then defined as
\begin{equation}
C_B(D) = \sup \{R|(R,D,B)~\textrm{is~achievable}\}.
\end{equation}\noindent

\begin{figure}[t]
    \centering
    \begin{overpic}[width=.46\textwidth]{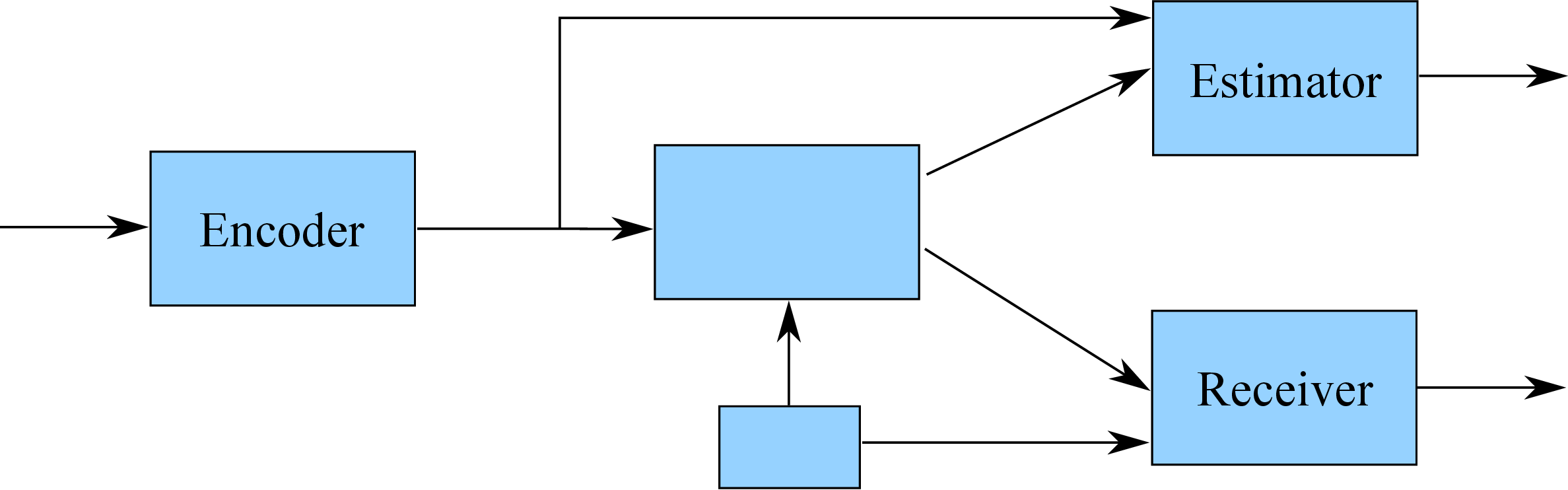}
        \put(42.2,16){\footnotesize $P_{\RM{Y}_{\rm s},\RM{Y}_{\rm c}|\RM{X},\RV{\eta}}$}
        \put(3,18){\footnotesize $\rv{W}$}    
        \put(29.5,17.5){\footnotesize $\RM{X}_i$}
        \put(61,25){\footnotesize $\RM{Y}_{{\rm s},i}$}
        \put(61,14){\footnotesize $\RM{Y}_{{\rm c},i}$}
        \put(93,7.5){\footnotesize $\hat{\rv{W}}$}
        \put(93.5,28){\footnotesize $\hat{\RV{\eta}}_i$}
        \put(48,2){\footnotesize $P_{\RV{\eta}}$}
        \put(62,4.5){\footnotesize $\RV{\eta}_i$}
    \end{overpic}
    \caption{The structure of a generic \ac{isac} system with memoryless channels, often considered in the capacity-distortion theory.}
    \label{fig:general_CD_model}
    \vspace{-3mm}
\end{figure}
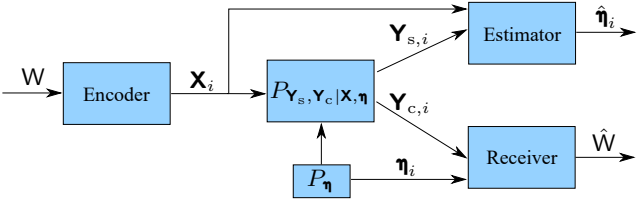

For the simplicity of discussion, let us consider memoryless channels,\footnote{It is possible that \ac{isac} channels have memory. However, the capacity-distortion problem of \ac{isac} channels with memory remains open in general, except for some special cases \cite{10144803}.} exemplified by \eqref{model_detection}, and illustrated in Fig.~\ref{fig:general_CD_model}. The effect of such channels can be expressed as
\begin{align}
&p_{\rv{W},\RM{Y}_{\rm c}^N,\RM{Y}_{\rm s}^N,\RM{X}^N,\RV{\eta}^N}(W,\M{Y}_{\rm s},\M{Y}_{\rm c},\M{X},\V{\eta})=p_{\rv{W}}(W) \nonumber\\ 
&\hspace{3mm}\times \prod_{i=1}^N p_{\RV{\eta}}(\V{\eta}_i)p_{\RM{X}|\rv{W}}(\M{X}_i|\rv{W})p_{\RM{Y}_{\rm s},\RM{Y}_{\rm c}|\RM{X},\RV{\eta}}(\M{Y}_{{\rm s},i},\M{Y}_{{\rm c},i}|\M{X}_i,\V{\eta}_i).
\end{align}
In particular, the channel $p_{\rv{W},\RM{Y}_{\rm c}^N,\RM{Y}_{\rm s}^N,\RM{X}^N,\RV{\eta}^N}(W,\M{Y}_{\rm s},\M{Y}_{\rm c},\M{X},\V{\eta})$ produces a communication output $\RM{Y}_{\rm c}$ and a sensing output $\RM{Y}_{\rm s}$, whereas their relationship with the channel input $\RM{X}$ can be different from the linear model \eqref{model_detection}. Moreover, the way that the sensing parameter $\RV{\eta}$ couples with the channel can also be different. For such channels, it is natural to consider per-block resource budgets and distortion metrics, given by
$$
\begin{aligned}
\mathbb{E}\big\{b\big(\RM{X}^N\big)\big\}&=\frac{1}{N}\sum_{n=1}^N\mathbb{E}\{b(\RM{X}_i)\},\\
\mathbb{E}\big\{d\big(\RV{\eta}^N,\hat{\RV{\eta}}^N\big)\big\}&=\frac{1}{N}\sum_{n=1}^N \mathbb{E}\{d(\RV{\eta}_i,\hat{\RV{\eta}}_i)\}.
\end{aligned}
$$
The conditions \eqref{capacity_distortion_cost} can then be concretized as
\begin{subequations}\label{capacity_distortion_dfsdmc}
\begin{align}
\limsup_{N\rightarrow \infty}~\frac{1}{N}\sum_{n=1}^N \mathbb{E}\{d(\RV{\eta}_i,\hat{\RV{\eta}}_i)\} &\leq D, \label{distortion_constraint}\\
\limsup_{N\rightarrow \infty}~\frac{1}{N}\sum_{n=1}^N\mathbb{E}\{b(\RM{X}_i)\} &\leq B, \\
\limsup_{N\rightarrow \infty}~P_{\rm e}^{(N)}&=0.
\end{align}
\end{subequations}

In order to gain insights from \eqref{capacity_distortion_dfsdmc}, observe that when the sensing distortion constraint \eqref{distortion_constraint} is absent, the capacity is given by the classical result of ``capacity-with-cost'' \cite{nit} 
\begin{equation}\label{capacity_with_cost}
C_B = \max_{p_{\RM{X}}(\M{X})\in\Set{P}_B}~~I(\RM{X};\RM{Y}_{\rm c}),
\end{equation}\noindent
where
$$
\Set{P}_B=\{p_{\RM{X}}(\M{X})|\mathbb{E}\{b(\RM{X})\}\leq B\}.
$$   
The result \eqref{capacity_with_cost} implies that the capacity in the presence of cost budget has a single-letter representation, which would extremely simplify further analysis. Furthermore, since the sensing channel is memoryless, the optimal estimator does not rely on historical observations, and hence the expected distortion can be written as a function of the channel input, i.e. $\mathbb{E}\{d(\RV{\eta}_i,\hat{\RV{\eta}}_i)|\RM{X}=\M{X}\}=c(\M{X})$. In light of this, the capacity-distortion function can now be viewed as a capacity with two costs, namely \cite{CD_TIT}
\begin{subequations}\label{capacity_distortion_as_two_costs}
\begin{align}
C_B(D)&=\sup_{p_{\RM{X}}(\M{X})\in\Set{P}_B\cap\Set{P}_D}~I(\RM{X};\RM{Y}_{\rm c}|\RV{\eta}), \label{capacity_distortion_same_channel}\\
\Set{P}_B&=\{p_{\RM{X}}(\M{X})|\mathbb{E}\{b(\RM{X})\}\leq B\}, \label{constraint_cost}\\
\Set{P}_D&=\{p_{\RM{X}}(\M{X})|\mathbb{E}\{c(\RM{X})\}\leq D\}. \label{constraint_distortion}
\end{align}
\end{subequations}

%Aiming at providing a single-letter representation for the capacity-distortion function, it is shown that \edit{for the \ac{sdmcdf} model}, the optimal sensing distortion is achievable by the simple letter-wise minimum expected posterior distortion estimator \cite{CD_TIT} taking the form of
%$$
%\gamma(\RM{X},\RM{Z}) = \left(\hat{\RV{\eta}}^\ast(\RM{X}_1,\RM{Z}_1),\dotsc,\hat{\RV{\eta}}^\ast(\RM{X}_N,\RM{Z}_N)\right),
%$$
%where
%$$
%\hat{\RV{\eta}}^\ast(\M{X},\M{Z}) = \mathop{\arg\min}_{\RV{\eta}^\prime}~\int d(\RV{\eta},\RV{\eta}^\prime)p_{\RV{\eta}|\RM{X},\RM{Z}}(\V{\eta}|\M{X},\M{Z}){\rm d}\V{\eta},
%$$
%and $p_{\RV{\eta}|\RM{X},\RM{Z}}(\V{\eta}|\M{X},\M{Z})$ is the \textit{a posteriori} distribution of $\RV{\eta}$ given the transmitted codeword $\RM{X}$ (known at the \ac{tx}) and the delayed feedback $\RM{Z}$. The intuition behind this result is that in memoryless channels, feedback from the past is not beneficial for the estimation of the current channel state. We may now see that the expected distortion also admits a single-letter representation
%\begin{equation}\label{sensing_single_letterization}
%c(\M{X}) = \mathbb{E}\big\{d\big(\RV{\eta},\hat{\RV{\eta}}^\ast(\RV{X},\RV{Z})\big)|\RM{X}=\M{X}\big\}.
%\end{equation}\noindent

Remarkably, the result \eqref{capacity_distortion_as_two_costs} suggests that the expected sensing distortion may be alternatively viewed as a kind of ``sensing-induced communication cost''. This understanding enables us to extend even further the range that the capacity-distortion theory is applicable, to scenarios in which the sensing performance cannot be described by a proper distortion function. For example, the \ac{crb} widely used as the performance metric of estimation problems is not dependent on any specific estimator $\hat{\RV{\eta}}$. Sometimes it does not even depend on the true value of the parameter $\RV{\eta}$. Therefore, \ac{crb} is not a distortion function, but it is related to the transmitted codeword $\RM{X}$ (as will be discussed later), and hence the capacity-distortion theory can still be applied in the generalized sense.

In this aforementioned scenario, we did not consider feedback. One may wonder whether designing the \ac{isac} codebook by relying on feedback can further enhance the \ac{snc} performance. To this end, the \ac{sdmcdf} model has been investigated \cite{CD_TIT}. The effect of an \ac{sdmcdf} can be expressed as \cite{cd}
\begin{align}
    &p_{\rv{W},\RM{Y}^N,\RM{Z}^N,\RM{X}^N,\RV{\eta}^N}(W,\M{Y},\M{Z},\M{X},\V{\eta})=p_{\rv{W}}(W)\nonumber \\
    &\hspace{1mm}\times\!\prod_{i=1}^{N}p_{\RV{\eta}}(\V{\eta}_i)p_{\RM{X}|\rv{W},\RM{Z}}(\M{X}_i|\rv{W},\!\M{Z}_{i-1})p_{\RM{Y},\RM{Z}|\RM{X},\RV{\eta}}(\M{Y}_i,\!\M{Z}_i|\M{X}_i,\!\V{\eta}_i)
\end{align}
where both sensing and communication rely on the same channel, which produces the output $\RM{Y}^N$. For sensing tasks, the channel input $\RM{X}_i$ at the $i$-th channel use cannot be designed based on the real-time channel output $\RM{Y}_i$ or the parameter value $\RV{\eta}_i$, since the design process has to be causal. Rather, one can only rely on a \emph{delayed feedback} $\RM{Z}_{i-1}$, which may be a function of the channel output $\RM{Y}_i$. 

It turns out that, for \ac{sdmcdf} models, \eqref{capacity_distortion_as_two_costs} also applies, with the slight modification that $\RM{Y}_{\rm c}$ is replaced with $\RM{Y}$. To see this, first note that feedback does not improve the capacity of memoryless channels. For the sensing performance, it has been shown that the optimal sensing distortion is achievable by the simple letter-wise minimum expected posterior distortion estimator \cite{CD_TIT} taking the form of
$$
\gamma(\RM{X},\RM{Z}) = \left(\hat{\RV{\eta}}^\ast(\RM{X}_1,\RM{Z}_1),\dotsc,\hat{\RV{\eta}}^\ast(\RM{X}_N,\RM{Z}_N)\right),
$$
where
$$
\hat{\RV{\eta}}^\ast(\M{X},\M{Z}) = \mathop{\arg\min}_{\RV{\eta}^\prime}~\int d(\RV{\eta},\RV{\eta}^\prime)p_{\RV{\eta}|\RM{X},\RM{Z}}(\V{\eta}|\M{X},\M{Z}){\rm d}\V{\eta}.
$$
We may now see that the expected distortion also admits a single-letter representation
\begin{equation}\label{sensing_single_letterization}
c(\M{X}) = \mathbb{E}\big\{d\big(\RV{\eta},\hat{\RV{\eta}}^\ast(\RV{X},\RV{Z})\big)|\RM{X}=\M{X}\big\},
\end{equation}\noindent
and thus \eqref{capacity_distortion_as_two_costs} still applies. 

From the aforementioned discussion, we may conclude that a key condition for \eqref{capacity_distortion_as_two_costs} to hold is the memorylessness of channels. Indeed, for channels with memory, even the capacity itself remains open. Furthermore, for such channels, online and offline estimators may have very different performances. These problems deserve further investigation.

\subsection{Computing the Capacity-Distortion Boundary}
From a pure theorist's perspective, the result \eqref{capacity_distortion_as_two_costs} is a complete information characterization (as opposed to the operational definition \eqref{capacity_distortion_dfsdmc}) of the capacity-distortion region. But wait! Remember that our expectation on the capacity-distortion theory in the first place was to help understand the nature of the \ac{snc} tradeoff, hopefully beyond the \ac{st}. However, we cannot even dig \ac{st} itself out of \eqref{capacity_distortion_as_two_costs}, not to mention any further insight. 

To serve our purpose, a possible approach is to compute explicitly the capacity-distortion functions for some representative scenarios, in the hope of obtaining useful intuitions. It turns out that the renowned \ac{ba} algorithm \cite{1054855} originally proposed for computing the unconstrained capacity can be applied to compute capacity-distortion functions with some modifications. To elaborate, given an initialization of the trail distribution $r(\M{X})$ for $p_{\RM{X}}(\M{X})$, the original \ac{ba} algorithm solving $C = \max_{p_{\RM{X}}(\M{X})}~I(\RM{X};\RM{Y})$ is a fixed-point iteration repeating the following two steps in each round:
\begin{enumerate}
    \item Update the trail distribution $q(\M{X}|\M{Y})$ for the \textit{a posteriori} distribution $p_{\RM{X}|\RM{Y}}(\M{X}|\M{Y})$ according to
    \begin{equation}\label{ba_step1}
        q(\M{X}|\M{Y})\leftarrow \frac{p_{\RM{X}}(\M{X})p_{\RM{Y}|\RM{X}}(\M{Y}|\M{X})}{\int r(\M{X})p_{\RM{Y}|\RM{X}}(\M{Y}|\M{X}) {\rm d}\M{X}};
    \end{equation}\noindent
    \item Update $r(\M{X})$ by 
    \begin{equation}\label{ba_step2}
        r(\M{X})\leftarrow \frac{e^{\int p_{\RM{Y}|\RM{X}}(\M{Y}|\M{X})\log q(\M{X}|\M{Y}){\rm d}\M{Y}}}{\int e^{\int p_{\RM{Y}|\RM{X}}(\M{Y}|\M{X})\log q(\M{X}|\M{Y}){\rm d}\M{Y}} {\rm d}\M{X}},
    \end{equation}\noindent
    which is equivalent to solving the following constrained entropy maximization problem:
    $$
    \begin{aligned}
        r(\M{X})\leftarrow &\mathop{\arg\max}_{p(\M{X})}  -\int p(\M{X})\log p(\M{X}) {\rm d}\M{X},\\
        &{\rm s.t. } ~p(\M{X})\geq 0~\forall \M{X},~\int p(\M{X}){\rm d}\M{X} = 1,\\
        &\mathbb{E}_{p(\M{X})}\Big\{\int p_{\RM{Y}|\RM{X}}(\M{Y}|\M{X})\log q(\M{X}|\M{Y}) {\rm d}\M{Y}\Big\}\geq t
    \end{aligned}
    $$
    for some constant $t$.
\end{enumerate}
Now that our objective function is the conditional mutual information $I(\RM{X};\RM{Y}|\RV{\eta})$, we shall replace \eqref{ba_step1} with
$$
q(\M{X}|\M{Y},\V{\eta})\leftarrow \frac{p_{\RM{X}}(\M{X})p_{\RM{Y}|\RM{X},\RV{\eta}}(\M{Y}|\M{X},\V{\eta})}{\int p_{\RM{X}}(\M{X})p_{\RM{Y}|\RM{X},\RV{\eta}}(\M{Y}|\M{X},\V{\eta}) {\rm d}\M{X}}.
$$
In addition, since two extra constraints \eqref{constraint_cost} and \eqref{constraint_distortion} are now enforced, we should replace \eqref{ba_step2} with
$$
\begin{aligned}
&r(\M{X})\leftarrow \\
&\hspace{3mm}\frac{e^{\int [p_{\RV{\eta}}(\V{\eta})p_{\RM{Y}|\RM{X},\RV{\eta}}(\M{Y}|\M{X},\V{\eta})\log q(\M{X}|\M{Y},\V{\eta})-\lambda b(\M{X})-\mu c(\M{X})]{\rm d}\V{\eta}{\rm d}\M{Y}}}{\int e^{\int [p_{\RV{\eta}}(\V{\eta})p_{\RM{Y}|\RM{X},\RV{\eta}}(\M{Y}|\M{X},\V{\eta})\log q(\M{X}|\M{Y},\V{\eta})-\lambda b(\M{X})-\mu c(\M{X})]{\rm d}\V{\eta}{\rm d}\M{Y}}{\rm d}\M{X}},
\end{aligned}
$$
where $\lambda$ and $\mu$ are the Lagrangian multipliers corresponding to the constraints \eqref{constraint_cost} and \eqref{constraint_distortion}, respectively. 

From the above discussions, we may get a vague sense that the sensing distortion requirements render the resulting \ac{isac} signal $\RM{X}$ ``less random'', as they impose constraints on the entropy maximization problem. Of course, this is not yet a rigorous statement with a well-defined meaning. To validate our intuition, let us consider the toy example of a real-valued, \ac{siso} Gaussian \ac{sdmcdf} channel model (with Rayleigh fading)
$$
\rv{Y}_i = \rv{\eta}_i\rv{X}_i+\rv{N}_i,
$$
where $\rv{\eta}_i$ and $\rv{N}_i$ are mutually independent zero-mean Gaussian variables with unit variance, while $\rv{X}_i$ is the \ac{isac} waveform satisfying the power constraint $\limsup_{N\rightarrow \infty} \frac{1}{N}\sum_{n=1}^N \mathbb{E}\{|\rv{X}_i|^2\}\leq B=10$dB. We consider the perfect feedback $\rv{Z}_i=\rv{Y}_i$ and the quadratic sensing distortion $d(\eta,\hat{\eta})=(\eta-\hat{\eta})^2$. In this scenario, the sensing-optimal estimator is the letter-wise \ac{mmse} estimator given by $\hat{\rv{\eta}}_{i,{\rm MMSE}}=\mathbb{E}\{\rv{\eta}_i|\rv{X}_i,\rv{Y}_i\}$, whose \ac{mse} can be calculated as
$$
\frac{1}{N}\sum_{i=1}^N\mathbb{E}\{(\rv{\eta}-\hat{\rv{\eta}}_{i,{\rm MMSE}})^2\}=\mathbb{E}\bigg\{\frac{1}{1+|\rv{X}|^2}\bigg\},
$$
where $\rv{X}$ is the single-letter representation of the channel input following the probability distribution $p_{\rv{X}}(X)$.

\begin{figure}
\centering
\subfloat[The capacity-distortion boundary]{
\centering
\includegraphics[width=.45\textwidth]{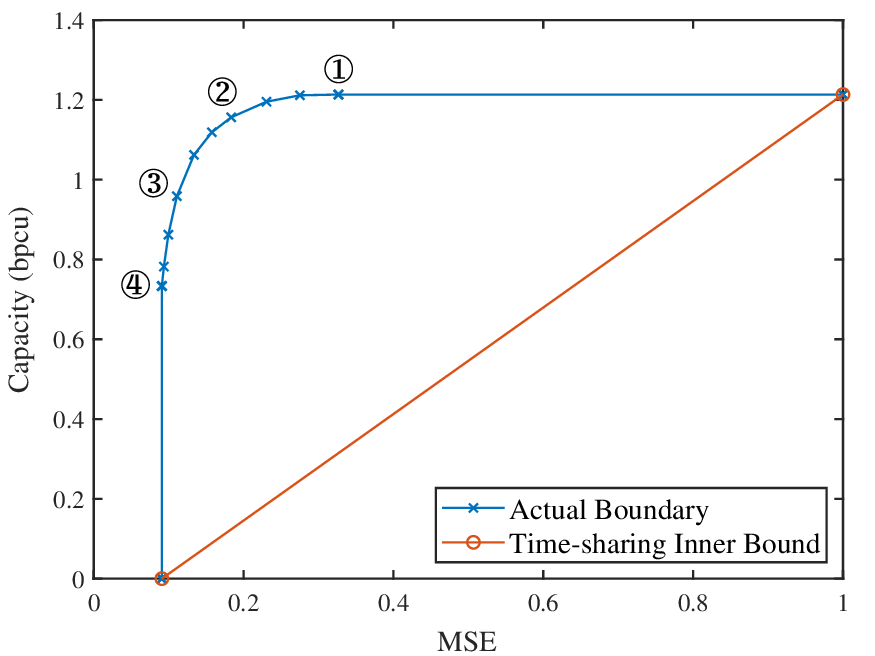}
}\\
\subfloat[$p_{\rv{X}}(X)$ at \textcircled{\tiny 1}]{
\includegraphics[width=.22\textwidth]{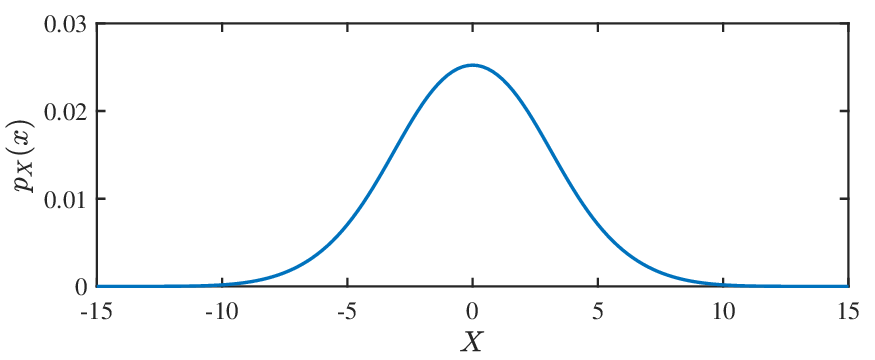}
}
\subfloat[$p_{\rv{X}}(X)$ at \textcircled{\tiny 2}]{
\includegraphics[width=.22\textwidth]{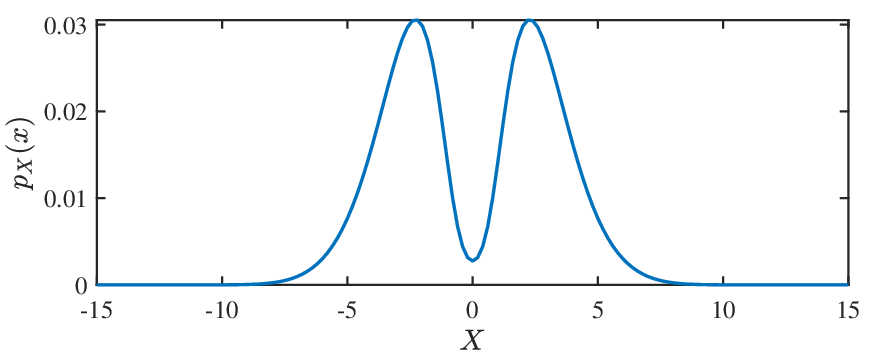}
}\\
\subfloat[$p_{\rv{X}}(X)$ at \textcircled{\tiny 3}]{
\includegraphics[width=.22\textwidth]{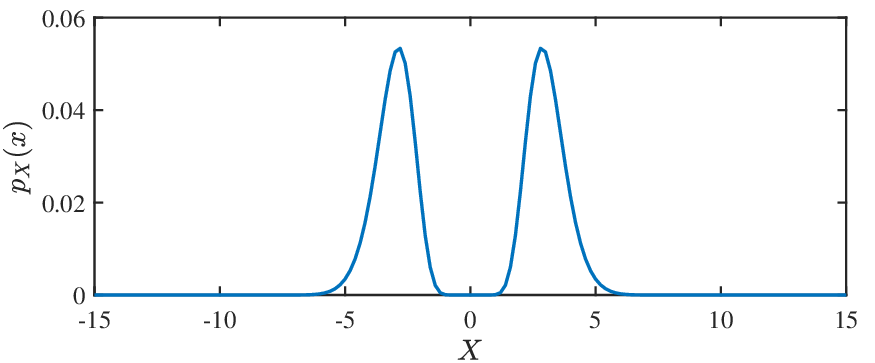}
}
\subfloat[$p_{\rv{X}}(X)$ at \textcircled{\tiny 4}]{
\includegraphics[width=.22\textwidth]{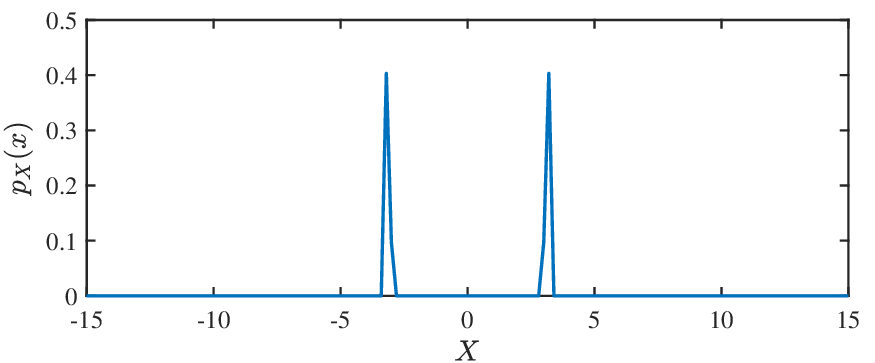}
}
\caption{The capacity-distortion boundary of the real-valued \ac{siso} Gaussian channel scenario with $B=10$dB, as well as the Pareto-optimal input distributions $p_{\rv{X}}(X)$ along the boundary.}
\label{fig:cd_siso_distribution}
\vspace{-3mm}
\end{figure}

Under the aforementioned formulation, the capacity-distortion boundary may be numerically computed using the modified \ac{ba} algorithm, as portrayed in Fig.~\ref{fig:cd_siso_distribution}. At the communication-optimal point $\textcircled{\footnotesize 1}$, the input distribution is Gaussian, and we have
$$
\begin{aligned}
C_B(D) = \frac{1}{2}\mathbb{E}\{\log_2(1+B|\rv{\eta}|^2)\} &\approx 1.213,\\
\mathbb{E}\{c(\rv{X})\}=\mathbb{E}\bigg\{\frac{1}{1+|\rv{X}|^2}\bigg\} &\approx 0.327.
\end{aligned}
$$
By contrast, at the sensing-optimal point $\textcircled{\footnotesize 4}$, the input distribution corresponds to \ac{bpsk} modulation, for which we have
$$
\begin{aligned}
C_B(D) &\approx 0.733,\\
\mathbb{E}\{c(\rv{X})\} =\frac{1}{1+B}&\approx 0.091.
\end{aligned}
$$
As we move along the capacity-distortion boundary from the communication-optimal point $\textcircled{\footnotesize 1}$ to the sensing-optimal point $\textcircled{\footnotesize 4}$, the corresponding input distribution $p_{\rv{X}}(X)$ exhibits a smooth transition from the Gaussian distribution to the \ac{bpsk} modulation, which agrees with our intuition that the \ac{isac} signal becomes less random as the sensing distortion requirement becomes more stringent. 

Note that there is not enough room to accommodate \ac{st} in the \ac{siso} scenario. Therefore, the \ac{snc} tradeoff in the aforementioned example follows solely from the difference in signal preference between communication and sensing tasks. Intuitively, communication tasks would favor signals with a higher degree of randomness, in order to pack more information into the signal. For example, the capacity for AWGN channels is achieved by using Gaussian-distributed signals, where the Gaussian distribution has the maximum entropy under power constraint.  By contrast, sensing tasks favor signals that are deterministic in some sense, in order to better distinguish the received signals (echoes) coupled with sensing parameters taking different values. Such a tradeoff is termed as \ac{drt} \cite{10147248}.

Naturally, we would then wonder whether the \ac{drt} exists in more general scenarios, what effect it would exhibit, and how it would interact with the \ac{st} when it exists. Unfortunately, although the modified \ac{ba} algorithm is universal in principle, it is hardly applicable to the general settings in a practical sense, due to its enormous computational complexity. To elaborate, the modified \ac{ba} algorithm relies on numerical integration, which could be computationally prohibitive when the dimensionality of the signals or the sensing parameters is high. For example, if the number of samples per dimension is $K$, the total number of samples $N_{\rm MC}$ that the widely used Monte Carlo integration method would require will be on the order of $N_{\rm MC} = O\big(K^{N_{\RM{Y}}+N_{\RM{X}}+N_{\RV{\eta}}}\big)$, where $N_{\RM{Y}}$, $N_{\RM{X}}$ and $N_{\RV{\eta}}$ are the dimensionalities of $\RM{Y}$, $\RM{X}$, and $\RV{\eta}$, respectively. The exponentially increasing sample complexity would easily surpass the capability of most computing devices, even for small-scale \ac{mimo} systems. Furthermore, the modified \ac{ba} algorithm cannot provide analytical solutions, which can be useful for practical system design.

Considering these difficulties, to push further our understanding of the \ac{snc} tradeoff, we might need to sacrifice the generality of the capacity-distortion theory to some degree. For example, we may try and find specific distortion functions (or non-distortion sensing-induced communication costs, as implied by \eqref{capacity_distortion_as_two_costs}), that are both easy to analyze and sufficiently general to reflect \ac{st} and \ac{drt}. In what follows, we shall consider one such example in detail.

\section{CRB-Rate Region}
We now focus our attention from the distortion measure $d(\RV{\eta},\hat{\RV{\eta}})$ to a specific sensing metric, namely, CRB for target parameter estimation. Unlike the MSE that specifically relies on the employed estimator, CRB serves as a globally lower bound for all unbiased estimators (satisfying the regularity condition), which usually leads to more tractable analytical expressions, and is achievable at high-SNR regimes \cite{kay1998fundamentals3}. Recalling the single-letterization of the expected distortion in (\ref{sensing_single_letterization}), one may treat the CRB as a cost function of the transmitted codeword $\RV{X}$, and consider the interplay between the CRB and communication rate as a special case of the C-D tradeoff. 

\subsection{Vector Gaussian System Model}
We commence by re-examining the model in (\ref{model_detection}), which may be extended to a more generic form as
\begin{subequations}\label{model_estimation}
\begin{align}
\RM{Y}_{{\rm c}} &= \RV{H}_{\rm c}\RM{X}+\RM{Z}_{{\rm c}},\label{comm_model} \\ 
\RM{Y}_{{\rm s}} &= \RV{H}_{\rm s}(\RV{\eta})\RM{X}+\RM{Z}_{{\rm s}}, \label{sensing_model} 
\end{align}
\end{subequations}
where the sensing channel $\RV{H}_{\rm s} \in \mathbb{C}^{{N_{\rm s}} \times M}$ is now defined as a deterministic, possibly nonlinear function of the sensing parameter $\RV{\eta} \in \mathbb{R}^K$, e.g., angular MIMO radar channel \cite{li2007mimo}. If not otherwise specified, the transmitted codeword $\RM{X} \in \mathbb{C}^{M\times T}$ will be referred to as an ISAC signal matrix in this section. This model may be viewed as a special case of the generic model shown in Fig.~\ref{fig:general_CD_model}. Following the preceding memoryless channel assumption, both the sensing parameter $\RV{\eta}$ and communication channel $\RV{H}_{\rm c} \in \mathbb{C}^{{N_{\rm c}} \times M}$ vary every $T$ symbols in an i.i.d. manner. The discussion on channels with memory (e.g., Markov channels) are designated as our future works. For convenience, we also assume that the ISAC Tx has perfect knowledge on $\RV{H}_{\rm c}$. Finally, $\RM{Z}_{{\rm c}}$ and $\RM{Z}_{{\rm s}}$ are zero-mean white Gaussian noise matrices with variances of $\sigma_{{\rm c}}^2$ and $\sigma_{{\rm s}}^2$, respectively.

\begin{figure}[t]
\centering
\begin{minipage}{.22\textwidth}
 \centering
\includegraphics[width=.95\textwidth]{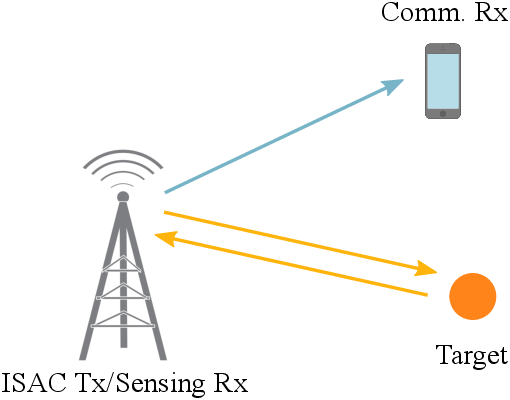}
\vspace{-1mm}
 \footnotesize (a) Monostatic sensing
\end{minipage}
\hspace{5mm}
\begin{minipage}{.22\textwidth}
\centering
\includegraphics[width=.95\textwidth]{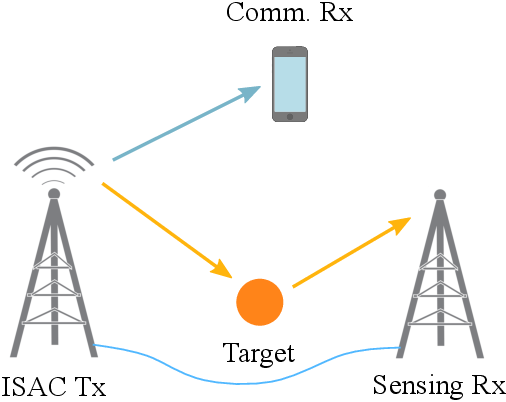}
  \footnotesize (b) Bistatic sensing
\end{minipage}
\caption{The \ac{isac} scenarios described in \eqref{model_estimation}, where the dual-functional waveform $\RV{X}$ is known to both the \ac{isac} \ac{tx} and sensing \acf{rx}.}
\label{fig:scenarios}
\vspace{-3mm}
\end{figure}

At this point, it would be worthwhile to concretize the separate channel models in \eqref{model_estimation} by illustrating the specific scenarios concerned in this section. As shown in Fig.~\ref{fig:scenarios}, the \ac{isac} \ac{tx} emits a dual-functional signal $\RV{X}$ to simultaneously communicate with a single communication \ac{rx} and sense one or more targets. The sensing \ac{rx} is either collocated with the \ac{tx} (monostatic sensing), or connected with the \ac{tx} with a wired link (bistatic sensing). In both cases, both the \ac{tx} and sensing \ac{rx} have perfect knowledge about the ISAC signal $\RV{X}$. Nevertheless, $\RV{X}$ is unknown to the communication \ac{rx} as it conveys useful information intended for the communication user. We are thus temped to model $\RV{X}\sim p_{\RM{X}}(\M{X})$ as a random matrix, whose realization is perfectly known at both the ISAC \ac{tx} and sensing \ac{rx}, but is unknown at the communication \ac{rx}.

\subsection{CRB with Random but Known Nuisance Parameters}
Since the communication performance of (\ref{model_estimation}a) can be directly measured by the mutual information $I(\RM{X};\RM{Y}_{\rm c}|\RV{H}_{\rm c})$, we are now in a position to rethink the sensing performance evaluation in ISAC systems. Indeed, in conventional radar systems, probing signals are typically deterministic, well-designed with good ambiguity properties. In ISAC systems, on the other hand, the transmit signal randomly varies from block to block given the communication data embedded in the signal. This imposes a unique challenge in defining the CRB, which now becomes a function of the random signal $\RV{X}$.

One possible approach would be to treat $\RV{X}$ as a nuisance parameter, and either consider it as a part of the unknown sensing parameter, or integrate it out of the likelihood function describing the observations, leading to classical hybrid or marginal CRB expressions, respectively. Nonetheless, neither of these methods grasps the fundamental feature of the ISAC system, that the random signal $\RV{X}$ is {\it{known}} to the sensing \ac{rx}, and is thus loose in general. To that end, we resort to a Miller-Chang type CRB \cite{mcb} by computing the CRB for a given instance of $\RV{X}$, and take the expectation over $\RV{X}$. For any weakly unbiased estimator of $\RV{\eta}$, the MSE is lower bounded by the Miller-Chang Bayesian CRB in the form of
\begin{equation}\label{MCB}
{\rm MSE}_{\RV{\eta}}(\hat{\RV{\eta}})\geq \mathbb{E}\left(\tr{\RM{J}_{\RV{\eta}|\RM{X}}^{-1}}\right),
\end{equation}\noindent
where the expectation is taken with respect to $\RM{X}$, and $\RM{J}_{\RV{\eta}|\RM{X}}$ denotes the Bayesian Fisher Information Matrix (BFIM) of $\RV{\eta}$ given by
\begin{equation}\label{BFIM}
\begin{aligned}
\RM{J}_{\RV{\eta}|\RM{X}}&\!:= \!\mathbb{E} \!\left\{\!\frac{\partial\ln p_{\RM{Y}_{\rm s}|\RM{X},\RV{\eta}}(\M{Y}_{\rm s}|\M{X},\V{\eta})}{\partial \V{\eta}}\frac{\partial\ln p_{\RM{Y}_{\rm s}|\RM{X},\RV{\eta}}(\M{Y}_{\rm s}|\M{X},\V{\eta})}{\partial \V{\eta}^{\rm T}} \bigg|\RM{X}\right\}\\
&\hspace{5mm}+\mathbb{E}\left\{\frac{\partial \ln p_{\RV{\eta}}(\V{\eta})}{\partial \V{\eta}}\frac{\partial \ln p_{\RV{\eta}}(\V{\eta})}{\partial \V{\eta}^{\rm T}}\right\}.
\end{aligned}
\end{equation}\noindent
More precisely, the BFIM $\RM{J}_{\RV{\eta}|\RM{X}}$ can be expressed as an affine map of the sample covariance matrix $\RM{R}_{\RM{X}}=T^{-1}\RM{X}\RM{X}^{\rm H}$ \cite{10147248}
\begin{equation}
\RM{J}_{\RV{\eta}|\RM{X}} =  \frac{T}{\sigma_{\rm s}^2}\M{\Phi}(\RM{R}_{\RM{X}}),
\end{equation}\noindent
where
\begin{equation}\label{phi_def}
\M{\Phi}(\M{A})=\sum_{i=1}^{r_1}\widetilde{\M{F}}_i\M{A}^{\rm T} \widetilde{\M{F}}_i^{\rm H}+\sum_{j=1}^{r_2}\widetilde{\M{G}}_j\M{A}\widetilde{\M{G}}_j^{\rm H} + \widetilde{\M{J}}_{\rm P},
\end{equation}\noindent
and $\widetilde{\M{J}}_{\rm P} = \sigma_{\rm s}^2T^{-1}\M{J}_{\rm P}$, with the term $\M{J}_{\rm P}$ being contributed by the prior distribution $p_{\RV{\eta}}(\V{\eta})$, i.e., the second term in \eqref{BFIM}. In particular, the matrices $\widetilde{\M{F}}_i$ and $\widetilde{\M{G}}_j$ are partitioned from the Jacobian matrix $\RM{F}:=\frac{\partial \operatorname{vec}(\RV{H}_{\rm s}^{\ast})}{\partial \RV{\eta}}$.

By noting \eqref{MCB}, it turns out that the Miller-Chang CRB is nothing but an equivalent ``expected sensing distortion" discussed in Sec. III-A (though it is not a real distortion measure), and may hence be viewed as a sensing-induced cost imposed on signaling resources. Although with high dimensionality, one may still deduce useful results on the CRB-rate tradeoff by exploiting the affine structure of the BFIM, as discussed in the sequel.

\begin{figure}[t]
    \centering
    \includegraphics[width=.45\textwidth]{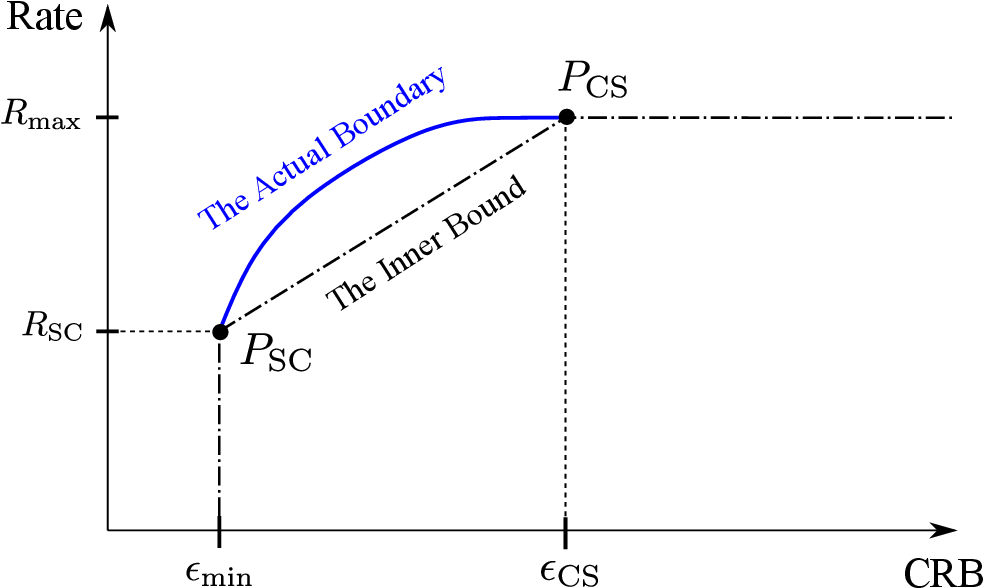}
    \caption{CRB-rate region, Pareto boundary, and time-sharing inner bound.}
    \label{fig:pentagon_inner_bound}
    \vspace{-3mm}
\end{figure}

\subsection{CRB-Rate Tradeoff}
The CRB-rate tradeoff can be characterized as the following Pareto optimization problem
\begin{subequations}\label{opt_problem}
\begin{align}
\min_{p_{\RM{X}}(\M{X})} &~~\alpha\mathbb{E}\left(\tr{\left[\M{\Phi}({\RM{R}}_{\RM{X}})\right]^{-1}}\right) - (1-\alpha)I(\RM{X};\RM{Y}_{\rm c}|\RV{H}_{\rm c}), \label{obj_function_CRB}\\
{\rm s.t.}&~~\mathbb{E}\left(\tr{{\RM{R}}_{\RM{X}}}\right)=P_{\rm T}, \label{opt_constraints}
\end{align}
\end{subequations}
where $\alpha \in \left[0,1\right]$ is a weighting factor controlling the priority of \ac{snc} performance. We highlight here that by moving the CRB from \eqref{obj_function_CRB} to \eqref{opt_constraints}, \eqref{opt_problem} may be equivalently recast as a constrained capacity characterization problem with two cost functions as in \eqref{capacity_distortion_as_two_costs}. 

Needless to say, fully depicting the Pareto boundary of the CRB-rate region would incur unaffordable computational overheads, as one has to numerically seek for optimal $p_{\RM{X}}(\M{X})$ by leveraging the modified B-A algorithm discussed in Sec. III-B. To reveal fundamental insights into the CRB-rate tradeoff, we are interested in the two corner points over the Pareto frontier as shown in Fig.~\ref{fig:pentagon_inner_bound}, namely, $P_{\rm CS}$, the minimum achievable sensing CRB constrained by the maximum communication capacity, and $P_{\rm SC}$, the maximum achievable rate constrained by the minimized CRB. It is obvious that the line segment linked to two points forms a time-sharing inner-bound. In what follows, we briefly characterize the \ac{snc} performance at the two points.

\subsection{$P_{\rm CS}$ Performance Characterization}
Let us first examine the point $P_{\rm CS}$. For point-to-point Gaussian channel, it is well-known that Gaussian distribution is the unique capacity-achieving input distribution (CAID) under an average power cost \cite{IT_Polyanskiy_Wu}, in which case \eqref{opt_problem} reduces to a capacity characterization problem with only power constraint by simply letting $\alpha = 0$. More specifically, at $P_{\rm CS}$, each column of $\RV{X}$ follows a circularly symmetric complex Gaussian distribution $\mathcal{CN}(\V{0},\widetilde{\M{R}}_{\RM{X}}^{\rm CS})$ in an i.i.d. manner, where the statistical covariance matrix $\widetilde{\M{R}}_{\RM{X}}^{\rm CS}$ is obtained by solving the following rate maximization problem
\begin{equation}\label{capacity_CS}
\begin{gathered}
  {R_{\rm{max}}} = \mathop {\max }\limits_{\widetilde{\M{R}}\succcurlyeq {\M{0}},\widetilde{\M{R}} = {\widetilde{\M{R}}^{\rm H}}} \mathbb{E}\left\{\log \left| {{\M{I}} + \sigma _{c}^{ - 2}{{\RV{H}}_{c}}\widetilde{\M{R}}{\RV{H}}_{c}^{\rm H}} \right|\right\}\hfill \\\;\;\;\;\;\;\;\;\;\;\;\;\;\;\;\;\;\;\;{\rm s.t.}\;\;\;\;\;\tr{\widetilde{\M{R}}} \leq {P_{\rm T}} \hfill \\
  \;\;\;\;\;\;\;\;\; = \mathbb{E} \left\{\log \left| {{\M{I}} + \sigma _{\rm c}^{ - 2}{{\RV{H}}_{\rm c}}\widetilde{\M{R}}_{\RM{X}}^{\rm CS} {\RV{H}}_{c}^{\rm H}} \right|\right\}. \hfill \\ 
\end{gathered}
\end{equation}\noindent
It is readily observed that the optimal solution of \eqref{capacity_CS} has the following eigenvalue decomposition structure
\begin{equation}
\widetilde{\M{R}}_{\RM{X}}^{\rm CS}= \M{U}_{\rm c}\M{\Lambda}_{\rm c}\M{U}_{\rm c}^{\rm H},
\end{equation}\noindent
where $\M{U}_{\rm c}$ contains the right singular vectors of ${{\RV{H}}_{c}}$, and $\M{\Lambda}_{\rm c}$ contains the optimal eigenvalues that can be attained from the water-filling method. Accordingly, the optimal ISAC signal structure at $P_{\rm CS}$ is
\begin{equation}\label{comms_optimal_x}
\RM{X}^{\rm CS} = \M{U}_{\rm c}\M{\Lambda}_{\rm c}^{\frac{1}{2}} \RM{D},
\end{equation}\noindent
where the entries of $\RM{D}$ i.i.d. subject to $\mathcal{CN}(0,1)$.

One would then raise the natural question: What is the sensing performance if a Gaussian signal is transmitted? From \eqref{MCB} to \eqref{phi_def}, it is obvious that the CRB is determined by the sample covariance matrix $\RM{R}_{\RM{X}} = T^{-1}\RM{X}\RM{X}^{\rm H}$ rather than the statistical covariance matrix $\widetilde{\M{R}}_{\RM{X}} = \mathbb{E}(\RM{R}_{\RM{X}})$. At $P_{\rm CS}$, since each column of $\RM{X}$ is i.i.d. Gaussian distributed, $\RM{R}_{\RM{X}}$ follows a complex Wishart distribution. Note the fact that ${{\text{tr}}\left( {{{\left[ {{\M{\Phi}} \left( {{\RM{R}_{\RM{X}}}} \right)} \right]}^{ - 1}}} \right)} $ is a convex function in $\RM{R}_{\RM{X}}$. Upon denoting the CRB at $P_{\rm CS}$ as $\epsilon_{\rm CS}$, and based on Jensen's inequality, we have
\begin{equation}\label{Jensen}
\begin{gathered}
  {\epsilon _{{\rm{CS}}}} \triangleq \frac{T}{\sigma_{\rm s}^2} {\mathbb{E}}\left\{ {{\text{tr}}\left( {{{\left[ {{\M{\Phi}} \left( {{\RM{R}_{\RM{X}}}} \right)} \right]}^{ - 1}}} \right)} \right\} \geq \frac{T}{\sigma_{\rm s}^2}{\text{tr}}\left\{ {{{\left( {{\M{\Phi}} \left[ {\mathbb{E}\left( {\RM{R}_{\RM{X}}} \right)} \right]} \right)}^{ - 1}}} \right\} \hfill \\
  \;\;\;\;\;\; = \frac{T}{\sigma_{\rm s}^2}{\text{tr}}\left\{ {{{\left[ {{\M{\Phi}} \left( \widetilde{\M{R}}_{\RM{X}}^{\rm CS} \right)} \right]}^{ - 1}}} \right\},\hfill \\ 
\end{gathered}
\end{equation}\noindent
which holds for arbitrarily distributed $\RM{R}_{\RM{X}}$. This suggests there is certain performance loss for sensing due to the Wishart distributed $\RM{R}_{\RM{X}}$, since the Jensen lower bound is attained when $\RM{R}_{\RM{X}} = \mathbb{E}(\RM{R}_{\RM{X}}) = \widetilde{\M{R}}_{\RM{X}}$, which holds for Wishart matrix only when $T\to\infty$.

A more non-trivial result in \cite{10147248} depicts the upper bound of ${\epsilon _{{\rm{CS}}}}$, given by 
\begin{equation}\label{sensing_dofloss}
{\mathbb{E}}\left\{ {{\text{tr}}\left( {{{\left[ {{\M{\Phi}} \left( {{\RM{R}_{\RM{X}}}} \right)} \right]}^{ - 1}}} \right)} \right\}\leq \frac{T\cdot \tr{\M{\Phi}(\widetilde{\M{R}}_{\RM{X}}^{\rm CS})^{-1}}}{T-\min\{K,\rv{M}_{\rm CS}\}},
\end{equation}\noindent
with $\rv{M}_{\rm CS}={\rm rank}(\widetilde{\M{R}}_{\RM{X}}^{\rm CS})$, which clearly indicates that the maximum sensing performance loss at $P_{\rm CS}$ is jointly determined by the number of sensing parameters $K$ and the rank\footnote{In high-SNR regime we have ${\rm rank}(\widetilde{\M{R}}_{\RM{X}}^{\rm CS}) = {\rm rank}(\RV{H}_{\rm c})$.} of $\widetilde{\M{R}}_{\RM{X}}^{\rm CS}$. Note again that when $T\to\infty$, the sensing performance is lossless since the upper bound converges to its lower counterpart.

\subsection{$P_{\rm SC}$ Performance Characterization}
The performance characterization at $P_{\rm SC}$ becomes more challenging compared to that of $P_{\rm CS}$, as the achieving strategy remains unknown in general. By denoting the achievable CRB at $P_{\rm SC}$ as $\epsilon_{\min}$, and using the Jensen's inequality again, we have
\begin{equation}\label{Jensen_SC}
\begin{gathered}
  \frac{T}{\sigma_{\rm s}^2} {\mathbb{E}}\left\{ {{\text{tr}}\left( {{{\left[ {{\M{\Phi}} \left( {{\RM{R}_{\RM{X}}}} \right)} \right]}^{ - 1}}} \right)} \right\} \geq \frac{T}{\sigma_{\rm s}^2}{\text{tr}}\left\{ {{{\left( {{\M{\Phi}} \left[ {\mathbb{E}\left( {\RM{R}_{\RM{X}}} \right)} \right]} \right)}^{ - 1}}} \right\} \hfill \\
  = \frac{T}{\sigma_{\rm s}^2}{\text{tr}}\left\{ {{{\left[ {{\M{\Phi}} \left( \widetilde{\M{R}}_{\RM{X}}^{\rm SC} \right)} \right]}^{ - 1}}} \right\} \triangleq \epsilon_{\min} \hfill \\ 
\end{gathered}
\end{equation}\noindent
holds for any complex semidefinite matrix ${\RM{R}_{\RM{X}}}$ satisfying the average power constraint, where $\widetilde{\M{R}}_{\RM{X}}^{\rm SC}$ is the solution of the deterministic CRB minimization problem
\begin{equation}\label{BCRB_opt}
{{\widetilde{\M{R}}_{\RM{X}}^{\rm{SC}}}} = \mathop {\arg \min }\limits_{{\widetilde{\M{R}}} \succcurlyeq {\mathbf{0}},\;{\widetilde{\M{R}}} = {{\widetilde{\M{R}}}^{\rm H}}} \;{\text{tr}}\left\{\left[ {{{\M{\Phi }}}\left( {\widetilde{\M{R}}} \right)} \right]^{ - 1}\right\}\;\;\operatorname{s.t.}\;{\text{tr}}\left( {\widetilde{\M{R}}} \right) \leq {P_{\rm T}}.
\end{equation}\noindent
While problem \eqref{BCRB_opt} is a convex semidefinite programming (SDP), it is not strictly convex. As a result, the optimal solution is not unique. In fact, all the solutions of an SDP belongs to the subspace spanned by the maximum-rank solution, hence may be parameterized as
\begin{equation}\label{sensing_optimal_rx}
\widetilde{\M{R}}_{\RM{X}}^{\rm{SC}} = \M{U}_{\rm s}\M{\Lambda}_{\rm s}\M{U}_{\rm s}^{\rm H},
\end{equation}\noindent
where $\M{U}_{\rm s}$ consists of the eigenvectors of the maximum-rank sensing-optimal solution corresponding to the non-zero eigenvalues, while $\M{\Lambda}_{\rm s}$ is a positive semidefinite Hermitian matrix.

Provably, \eqref{BCRB_opt} admits an unique solution in most situations \cite{10147248}, where the equality in \eqref{Jensen_SC} holds iff
\begin{equation}
    \RM{R}_{\RM{X}} = \mathbb{E}(\RM{R}_{\RM{X}}) = \widetilde{\M{R}}_{\RM{X}}^{\rm{SC}}.
\end{equation}\noindent
That is, the sample covariance matrix $\RM{R}_{\RM{X}}$ becomes a deterministic matrix when the global minimum of the CRB $\epsilon_{\min}$ is attained. One may then wonder whether there is any communication DoF left in the ISAC signal at $P_{\rm SC}$. The answer is non-trivially yes. This is because a deterministic $\RM{R}_{\RM{X}}$ does not necessarily imply a deterministic $\RV{X}$. The latter may still be a random signal conveying information, given by
\begin{equation}\label{sensing_optimal_x}
\RM{X}^{\rm SC} = \sqrt{T}(\widetilde{\M{R}}_{\RM{X}}^{\rm{SC}})^{\frac{1}{2}}\RM{Q} = \sqrt{T}\M{U}_{\rm s}\M{\Lambda}_{\rm s}^{\frac{1}{2}}\RM{Q},
\end{equation}\noindent
where $\RM{Q} \in \mathbb{C}^{\rv{M}_{\rm SC}\times T}$ is a random semi-unitary matrix such that $\RM{Q}\RM{Q}^{\rm H} = \M{I}$, where $\rv{M}_{\rm SC} = {\rm rank}(\widetilde{\M{R}}_{\RM{X}}^{\rm{SC}})$. Since $(\widetilde{\M{R}}_{\RM{X}}^{\rm{SC}})^{\frac{1}{2}}$ is deterministic, the communication DoFs at $P_{\rm SC}$ are only contributed by the randomness of $\RM{Q}$.
%\footnote{\edit{Defined as $\nu_{\rm c} = \lim_{\gamma \rightarrow \infty} \frac{R(\gamma)}{\log(1+\gamma)}$ where $R(\gamma)$ denotes the maximum achievable rate}

We are now ready to characterize the achievable communication rate at $P_{\rm SC}$. That is, seeking for the optimal distribution $p_{\RM{Q}}({\M{Q}})$ over the set of all $\rv{M}_{\rm SC}\times T$ semi-unitary matrices, namely, the Stiefel manifold $\mathcal{S}(T, \rv{M}_{\rm SC})$, such that the mutual information $I(\RM{Q};\RV{Y}_{\rm c}|\RV{H}_{\rm c})$ is maximized. In the high-SNR regime, this is equivalent to solving the sphere packing problem over the Stiefel manifold, where the optimal $p_{\RM{X}}(\M{X})$ is uniform distribution, leading to the following asymptotic achievable rate \cite{10147248}
\begin{equation}\label{comm_dof_loss}
R_{\rm SC} = \mathbb{E}\Big\{\Big(1-\frac{\rv{M}_{\rm SC}}{2T}\Big)\log |\sigma_{\rm c}^{-2}\RM{H}_{\rm c}\widetilde{\M{R}}_{\RM{X}}^{\rm SC}\RM{H}_{\rm c}^{\rm H}|+\rv{c}_0\Big\} + O(\sigma_{\rm c}^2),
\end{equation}\noindent
where 
\begin{equation}\label{c0_r1}
\rv{c}_0=\frac{\rv{M}_{\rm SC}}{T}\Big[\Big(T\!-\!\frac{\rv{M}_{\rm SC}}{2}\Big)\log\frac{T}{e}\!-\!\log \Gamma(T)\!+\!\log(2\sqrt{\pi})\Big]
\end{equation}\noindent
converges to zero as $T\rightarrow \infty$.

Observe immediately that when $T\rightarrow \infty$, the communication DoFs are lossless, since even a Gaussian matrix would have asymptotically orthogonal rows with the increase of $T$, making it asymptotically equivalent to a semi-unitary matrix.

\begin{figure}[t]
    \centering
    \includegraphics[width=.45\textwidth]{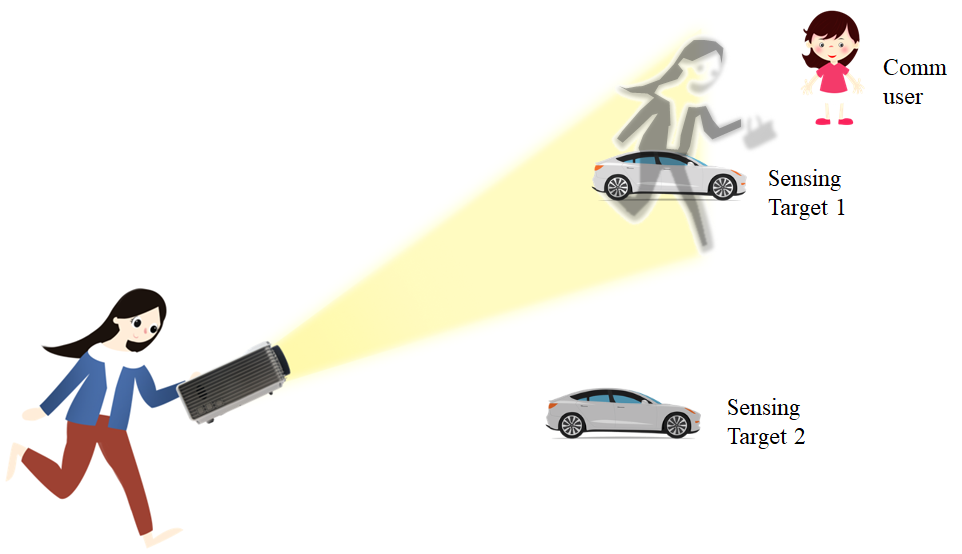}
    \caption{Graphical illustration of the projector metaphor.}
    \label{fig:projector_metaphor_illustration}
    \vspace{-3mm}
\end{figure}

\subsection{The Two-fold \ac{snc} Tradeoff: A Projector Metaphor}
The above results clearly demonstrate the effect of both ST and DRT in an ISAC system. By comparing the left-most parts of $\RM{X}^{\rm CS}$ and $\RM{X}^{\rm SC}$, we see that the communication- and sensing-optimal ISAC signals should be aligned to $\M{U}_{\rm c}$ and $\M{U}_{\rm s}$, respectively, which may be regarded as the orthogonal bases of communication and sensing subspaces. Then the ST is nothing but to flow the signal power from $\operatorname{span}(\M{U}_{\rm c})$ to $\operatorname{span}(\M{U}_{\rm s})$, perfectly fitting the picture of the ``ISAC torch" metaphor. More interestingly, by comparing the right-most parts of $\RM{X}^{\rm CS}$ and $\RM{X}^{\rm SC}$, e.g., $\RM{D}$ and $\RM{Q}$, we see that communication- and sensing-optimal signals adopt Gaussian and semi-unitary codebooks, respectively, which again reflects the DRT. That is, when the ISAC system moves along the Pareto frontier from $P_{\rm CS}$ to $P_{\rm SC}$, $p_{\RM{X}}(\M{X})$ gradually changes from Gaussian to less random distribution, and eventually becomes uniform distribution over semi-unitary matrices. In that sense, the DRT considered in Fig.~\ref{fig:cd_siso_distribution} is simply a one-dimensional special case, since the semi-unitary matrix reduces to constant-modulus signals in its scalar form, namely, the BPSK modulation in Fig.~\ref{fig:cd_siso_distribution}(e). 

In addition to the tradeoff between input distributions, the DRT may also be observed in the attainable communication and sensing DoF at the two corner points. At $P_{\rm CS}$, the communication subsystem apparently acquires the full DoF of the Gaussian channel, namely, $\rv{M}_{\rm CS}$, which is the rank of $\widetilde{\M{R}}_{\RM{X}}^{\rm CS}$. As shown in 
\eqref{sensing_dofloss}, due to the Gaussian signaling, the sensing subsystem suffers from the DoF loss, or, equivalently, the reduction in the number of individual observations of the target, which is up to $\min\{K,\rv{M}_{\rm CS}\}$. At $P_{\rm SC}$, the sensing subsystem attains the full DoF of $T$ thanks to the deterministic signal sample covariance matrix. In contrast, the semi-unitary signaling incurs a communication DoF loss of $\frac{\rv{M}_{\rm SC}^2}{2T}$, as indicated by \eqref{comm_dof_loss}.

To achieve the \ac{snc} performance tradeoff, it is critical to determine the steering direction of the ISAC signal. More importantly, what kind of codebook is transmitted through that direction also matters. With the above understanding, we may now correct the ``ISAC torch" metaphor with a more comprehensive picture, i.e., the ``ISAC projector". As shown in Fig.~\ref{fig:projector_metaphor_illustration}: A child (the ISAC \ac{tx}) holds a projector, and wishes to simultaneously illuminate a target (sensing) while sending an image to a receiver (communication). To form an image, the brightness of each pixel may be used to convey information. Nevertheless, those dark pixels result in imperfect illumination of the target.

\section{\ac{drt} and \ac{st} in Practical \ac{isac} Systems}
\subsection{DRT: Sensing with Random Signals}
The lessons learned from the DRT inspired us to rethink the practical design philosophy of ISAC systems. In particular, one has to take the randomness of communication data into account while conceiving a sensing strategy, which is a unique challenge emerged in the context of ISAC. In this subsection, we investigate a novel precoding design for sensing with random signals.

Let us consider again the sensing model \eqref{sensing_model}, with the sensing parameters being the entries of the channel matrix $\RV{H}_{\rm s}$, i.e., $\RV{\eta} = \operatorname{vec}(\RV{H}_{\rm s})$. For a given instance of $\RV{X}$, the linear minimum MSE (LMMSE) estimator reads
\begin{equation}
\hat{\RV{H}}_{\rm{LMMSE}}=\RV{Y}_{\rm s}\left(\RV{X}^{\rm H} \M{R}_{\RM{H}}\RV{X}+\sigma_{\rm s}^2 N_{\rm s} \M{I}\right)^{-1} \RV{X}^{\rm H} \M{R}_{\RM{H}},
\end{equation}
where $\M{R}_{\RM{H}} = \mathbb{E}(\RV{H}_{\rm s}^{\rm H}\RV{H}_{\rm s})$ represents the channel correlation matrix. The resulting estimation error is given as
\begin{equation}\label{LMMSE_error}
    \xi_{\RV{H}_{\rm s}|\RM{X}} = \operatorname{tr}\Big[\Big(\M{R}_{\RM{H}}^{-1} + \frac{1}{\sigma_{\rm s}^2 N_{\rm s}} \RV{X}\RV{X}^{\rm H}\Big)^{-1}\Big].
\end{equation}
Once again, the estimation error relies on the instantaneous realization of the random ISAC signal $\RV{X}$. To depict the average sensing performance, we take the expectation of \eqref{LMMSE_error} over $\RV{X}$, yielding  
\begin{equation}\label{ELMMSE_error}
    \xi_{\RV{H}_{\rm s}} = \mathbb{E}\left\{\operatorname{tr}\Big[\Big(\M{R}_{\RM{H}}^{-1} + \frac{1}{\sigma_{\rm s}^2 N_{\rm s}} \RV{X}\RV{X}^{\rm H}\Big)^{-1}\Big]\right\}.
\end{equation}
We refer the term \eqref{ELMMSE_error} to as the ergodic LMMSE (E-LMMSE) \cite{sensing_random_signals}, as it may be understood as a time average over different realizations of $\RV{X}$. Obviously, it is lower-bounded by the Miller-Chang CRB in \eqref{MCB}. 

We now investigate a specific form of the ISAC signal by letting $\RV{X} = \M{W}\RV{S}$, where $\M{W} \in \mathbb{C}^{M\times M}$ is a precoding matrix, and $\RV{S} = \left[{\RV{s}_1},{\RV{s}_2},\ldots,{\RV{s}_T}\right] \in \mathbb{C}^{M\times T}$ contains column-wise i.i.d. data symbols, satisfying $\mathbb{E}(\RV{s}_i) = \M{0}$ and $\mathbb{E}({\RV{s}_i}\RV{s}_i^{\rm H}) = \M{I}$. A fundamental question to ask is: What is the optimal precoder $\M{W}$ that minimizes the E-LMMSE? In classical MIMO radar waveform design, strictly orthogonal signals are typically employed, namely, $\frac{1}{T}\RV{S}\RV{S}^{\rm H} = \M{I}$, where $\RV{S}$ is a semi-unitary matrix, corresponding to the $P_{\rm SC}$ point discussed in Sec. IV. In such a case, it is known that the LMMSE-optimal precoder has a water-filling structure given by \cite{MIMO_radar_MMSE_MI}
\begin{equation}\label{water_filling}
    \M{W}_{\mathsf{WF}} = \sqrt{\frac{\sigma_{\rm s}^2{N_{\rm s}}}{T}} \M{Q}\Big[ \Big(\mu_0 \M{I} - \M{\Lambda}^{-1}\Big)^{+} \Big]^{\frac{1}{2}},
\end{equation}
where $\M{Q}$ and $\M{\Lambda}$ contain the eigenvectors and eigenvalues of $\M{R}_{\RM{H}}$, respectively, and $\mu_0$ is a constant meeting the power constraint $\|\M{W}_{\mathsf{WF}}\|_F^2 = P_{\rm T}$. However, in the ISAC scenario, the water-filling solution \eqref{water_filling} may not be optimal due to the randomness in $\RV{S}$. Evidently, applying the Jensen's inequality to \eqref{ELMMSE_error} yields
\begin{align}\label{Jensen_LMMSE}
    \xi_{\RV{H}_{\rm s}} & =  \mathbb{E}\left\{\operatorname{tr}\Big[\Big(\M{R}_{\RM{H}}^{-1} + \frac{1}{\sigma_{\rm s}^2 N_{\rm s}} \M{W}\RV{S}\RV{S}^{\rm H}\M{W}^{\rm H}\Big)^{-1}\Big]\right\}\nonumber \\
    & \overset{(a)}{\geq} \mathrm{tr}\Big[\Big(\bm{R}_H^{-1} + \frac{1}{\sigma_{\rm s}^2 N_{\rm s}} \mathbb{E}_{\bm{S}} \Big\{\M{W}\RV{S}\RV{S}^{\rm H}\M{W}^{\rm H}\Big\}\Big)^{-1}\Big]\nonumber \\
    &\overset{(b)}{=}\mathrm{tr}\Big[\Big(\bm{R}_H^{-1} + \frac{T}{\sigma_{\rm s}^2 N_{\rm s}} \M{W}\M{W}^H\Big)^{-1}\Big],
\end{align}
where (a) is due to the convexity of $\xi_{\RV{H}_{\rm s}|\RM{X}}$ with respect to $\RV{S}\RV{S}^{\rm H}$, and (b) holds from the fact $\mathbb{E}(\RV{S}\RV{S}^{\rm H}) = T\M{I}$. The equality in (a) holds only asymptotically\footnote{To the best of our knowledge, most existing ISAC precoding literature overlooked the data randomness by assuming $\frac{1}{T}\RV{S}\RV{S}^{\rm H} \approx \M{I}$.} for $\frac{T}{M} \to \infty$, since $\frac{1}{T}\RV{S}\RV{S}^{\rm H} \neq \M{I}$ due to the i.i.d. assumption in the columns of $\RV{S}$. 

\begin{figure}
\centering
\subfloat[$M = 64, N_{\rm s} = 32$]{
\centering
\includegraphics[width=.45\textwidth]{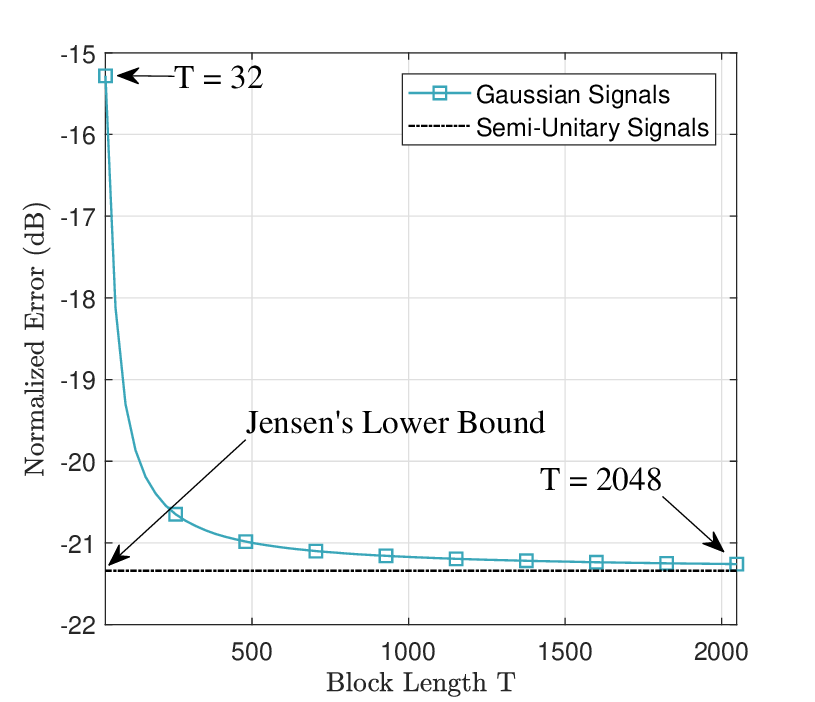}
\label{fig:Jensen_bound}
} \\
\subfloat[$M = 64, N_{\rm s} = 32, T = 32$]{
\centering
\includegraphics[width=.45\textwidth]{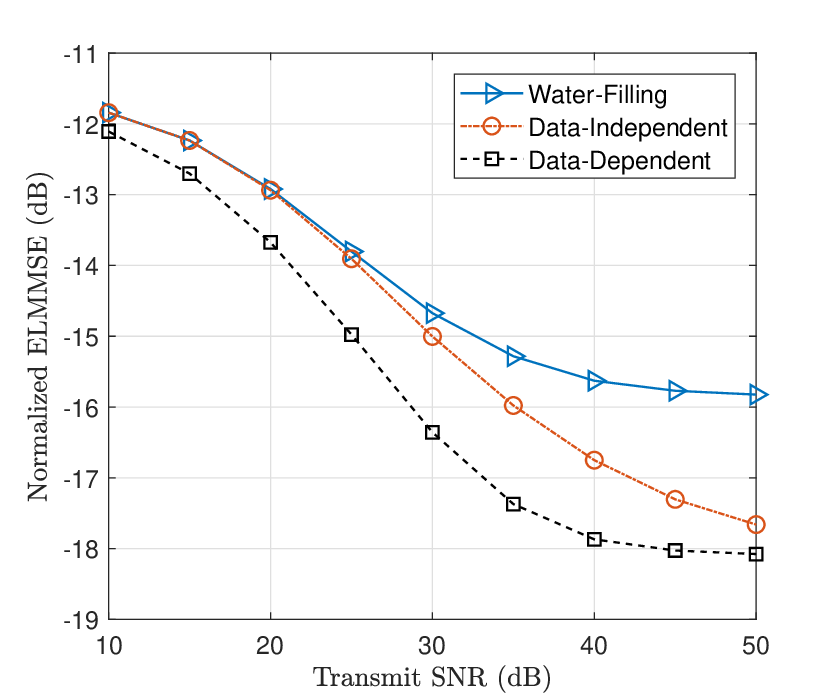}
\label{fig:LMMSE_L32Nt64}
}
\caption{Sensing with random signals. (a). Tightness of the Jensen bound with an increasing block length. (b). Estimation performance of the DRT-aware precoding designs.}
\label{fig:random_sensing}
\vspace{-3mm}
\end{figure}

It turns out that the water-filling solution \eqref{water_filling} minimizes the Jensen lower bound in \eqref{Jensen_LMMSE}, rather than $\xi_{\RV{H}_{\rm s}}$ itself. Specifically, when $T$ is comparable to $M$, and when non-unitary codebooks, e.g., Gaussian codebook (where $\RV{S}\RV{S}^{\rm H}$ follows Wishart distribution), are employed, the orthogonality of $\RV{S}$ breaks and the Jensen bound is no longer tight. To see this, we show an example in Fig.~\ref{fig:Jensen_bound} with $M = 64, N_{\rm s} = 32$, where the water-filling precoder \eqref{water_filling} is applied to both Gaussian distributed and semi-unitary data matrix. The Jensen bound (semi-unitary sensing performance) is attained when $T \geq 2048$. To account for the random nature of the ISAC signal, one needs to develop novel precoders that directly minimize the E-LMMSE, rather than its Jensen lower bound. Here we briefly introduce two possible methodologies, namely, data-dependent and data-independent designs, both of which yield better performance than that of the water-filling precoder \cite{sensing_random_signals}.
\subsubsection{Data-Dependent Precoding}
Despite its randomness, the fact that $\RV{S}$ is known at the ISAC Tx enables a precoding design based on given instances of $\RV{S}$. Let us denote the $n$th realization of $\RV{S}$ as $\M{S}_n$, and the to-be-designed precoder as $\M{W}_n$. One may then  directly minimize $\xi_{\RV{H}_{\rm s}|\RM{X}}$ based on the known $\M{S}_n$ by solving the following problem
\begin{equation}\label{data_dependent}
   \min_{\|\M{W}_{n}\|_F^2 = P_{\rm T}} \operatorname{tr}\Big[\Big(\M{R}_{\RM{H}}^{-1} + \frac{1}{\sigma_{\rm s}^2 N_{\rm s}} {\M{W}_n}{\M{S}_n}{\M{S}_n^{\rm H}}{\M{W}_n^{\rm H}}\Big)^{-1}\Big]. 
\end{equation}
While problem \eqref{data_dependent} is non-convex, it is provable that it admits an optimal closed-form solution \cite{5467189}. Consequently, one may minimize the E-LMMSE $\xi_{\RV{H}_{\rm s}}$ by minimizing every instance of $\xi_{\RV{H}_{\rm s}|\RM{X}}$.

\subsubsection{Data-Independent Precoding}
The optimality of the data-dependent precoder is achieved at the price of high complexity, as one has to solve for $\M{W}_n$ for every instance $\M{S}_n$. To ease the computational burden, an alternative option would be to conceive a data-independent precoder, where a single $\M{W}$ is leveraged for every instance of $\RV{S}$, leading to the following stochastic optimization problem
\begin{equation}\label{data_independent}
    \min_{\|\M{W}_{n}\|_F^2 = P_{\rm T}} \mathbb{E}\left\{\operatorname{tr}\Big[\Big(\M{R}_{\RM{H}}^{-1} + \frac{1}{\sigma_{\rm s}^2 N_{\rm s}} \M{W}\RV{S}\RV{S}^{\rm H}\M{W}^{\rm H}\Big)^{-1}\Big]\right\}, 
\end{equation}
which can be solved via stochastic gradient descent (SGD) algorithm in an offline manner, where massive training data samples may be locally generated based on the adopted communication codebook.

To validate the performance of the proposed sensing precoding design for random signals, we show their corresponding estimation errors in Fig.~\ref{fig:LMMSE_L32Nt64} with $M = 64, T = 32$ and $N_{\rm s} = 32$, where a Gaussian codebook is employed again for generating $\RV{S}$. As expected, both precoding designs significantly outperform the classical water-filling approach \eqref{water_filling}. In particular, the computationally expensive data-dependent design achieves better average estimation performance (0.4-1.4 dB gain) compared to its data-independent counterpart, while the latter attains a favorable performance-complexity tradeoff. This provides strong evidence that the data randomness is non-negligible in ISAC signaling. To achieve the \ac{snc} performance boundary, DRT-aware ISAC precoding techniques are yet to be implemented in practical ISAC systems.

\subsection{Frequency-domain ST: Valuating Sensing Resources}
Let us now turn our attention to the \ac{st}. In previous sections, we have illustrated the \ac{st} using spatial-domain examples. In these examples, roughly speaking, the \ac{st} is manifested as the preference for direct-illuminating beams, which holds for both the communication user and the sensing target. By contrast, when we consider non-spatial scenarios, the \ac{st} might take a less intuitive form, but it would also be more interesting and non-trivial.

To see this, let us consider the example of ranging waveform design, characterized by the simple observation model
$$
y(t) = s(t-\tau) + n(t),
$$
where $y(t)$, $s(t)$ and $n(t)$ represent the received signal, the transmitted signal, and the noise with constant \ac{psd} $N_0$, respectively. The term $\tau = d/c$ denotes the propagation delay, with $c$ being the propagation speed, and $d$ being the distance to be estimated. For this model, the \ac{crb} reads
\begin{equation}\label{crb_ranging}
\mathbb{E}\{(d-\hat{d})^2\}\geq c^2(8\pi^2\beta^2{\rm SNR})^{-1},
\end{equation}
where $\beta=\int_{-\infty}^{\infty}f^2|S(f)|^2{\rm d}f/\int_{-\infty}^{\infty}|S(f)|^2{\rm d}f$ is referred to as the ``\ac{rms} bandwidth'' \cite{anliu_fundamantal}, while ${\rm SNR}=\frac{1}{N_0}\int_{-\infty}^{+\infty}s^2(t){\rm d}t$. 

What does \eqref{crb_ranging} imply? Upon assuming that the signal is constrained to reside in the frequency interval of $[0,f_{\rm high}]$, we find immediately that the \ac{crb}-optimal signal is in fact a sinusoidal signal with frequency $f_{\rm high}$, which maximizes the \ac{rms} bandwidth. The intuition behind this result is that, as long as the integer ambiguity can be resolved, ranging methods based on carrier phase sensing would yield the optimal performance, as has been recognized in the literature of \ac{gnss}-based positioning \cite{8998218}.

\begin{figure}
    \centering
    \includegraphics[width=.45\textwidth]{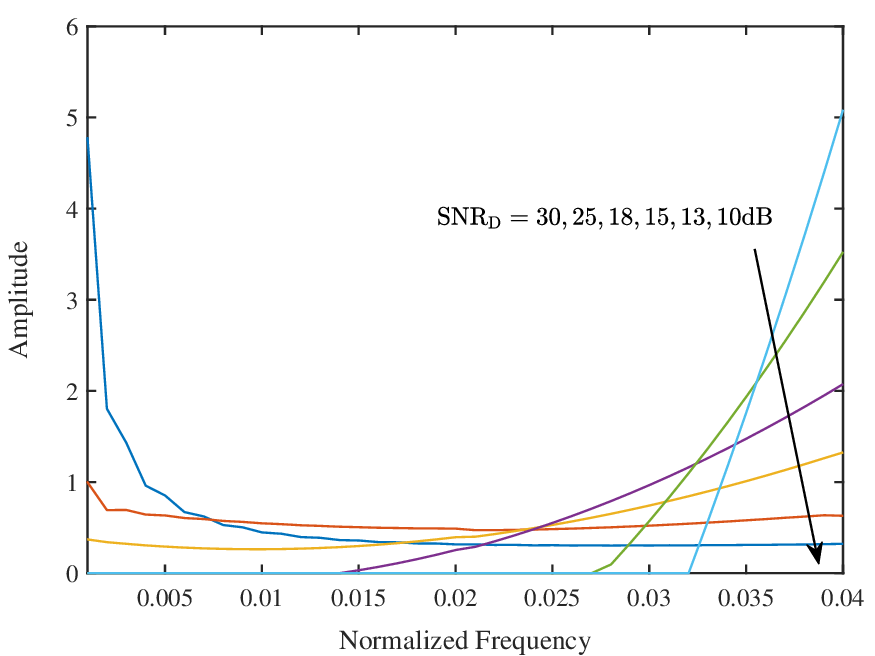}
    \caption{The \acp{psd} of \ac{zzb}-optimal waveforms at different \acp{snr}.}
    \label{fig:zzb_optimal_psd}
    \vspace{-3mm}
\end{figure}

Of course, if not supplemented by further information, the integer ambiguity of a single-tone signal can never be resolved. However, \ac{crb} is known to be unable to capture the ambiguity phenomenon. To this end, we may use the Ziv-Zakai bound (ZZB) given by \cite{ZZB_SPL}
\begin{equation}\label{zzb_ranging}
\mathbb{E}\left\{(d\!-\!\hat{d})^2\right\}\!\geq\! \int_0^{\epsilon_{\max}} x Q\Big(\sqrt{2^{-1}{\rm SNR}(1\!-\!\widetilde{R}(x))}\Big){\rm d}x,
\end{equation}
where $\epsilon_{\max}$ is the maximum possible ranging error, $Q(\cdot)$ denotes the Marcum Q-function, $\widetilde{R}(x)$ denotes the normalized \ac{acf} defined as $\widetilde{R}(x) = R(x/c)/R(0)$, with $R(\tau) = \int_{-\infty}^{\infty} s(t-\tau)s(t){\rm d}t$. Although it does not admit a closed-form expression, we may observe from \eqref{zzb_ranging} that the \ac{zzb} can reflect the ambiguity phenomenon: It is a decreasing function with respect to $(1-\widetilde{R}(x))$. Since the normalized \ac{acf} $\widetilde{R}(x)$ achieves its maximum at $\widetilde{R}(0)=1$, $(1-\widetilde{R}(x))$ can be viewed as a measure of the sidelobe level. Intuitively, given a fixed noise level, a higher sidelobe level would make it less distinguishable from the mainlobe, and hence cause larger errors.

With the aid of \ac{zzb}, we are now able to understand the behaviour of waveforms that achieve (near-) optimal sensing performance. Numerically computed \acp{psd} of \ac{zzb}-optimal waveforms (using the method in \cite{ZZB_SPL}) are plotted in Fig.~\ref{fig:zzb_optimal_psd}. Observe that as the total \ac{snr} increases, the \ac{zzb}-optimal waveform would refocus its power from the low-frequency band to the high-frequency band. The reason is that when the total \ac{snr} is sufficiently high, one may effectively resolve the ambiguity caused by relatively high sidelobes, and hence the power is focused on the high-frequency band. By contrast, when the total \ac{snr} is lower, one needs to lower the sidelobe level to fight against the ambiguity issue, which would inevitably widen the mainlobe, leading to low-frequency waveforms.\footnote{The curvature of the \ac{acf} mainlobe at $\tau=0$ is proportional to the \ac{rms} bandwidth \cite{anliu_fundamantal}. Therefore, wider mainlobes correspond to lower-frequency waveforms.}

\begin{figure*}
\centering
\subfloat[\scriptsize Communication Subspace]{
\begin{minipage}{.35\textwidth}
 \centering
\includegraphics[width=.99\textwidth]{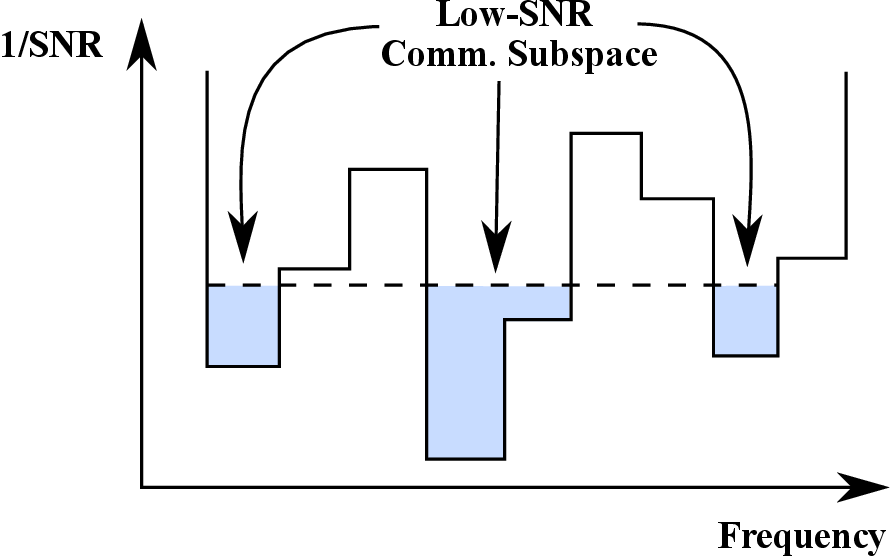}
\end{minipage}
\begin{minipage}{.35\textwidth}
\centering
\includegraphics[width=.99\textwidth]{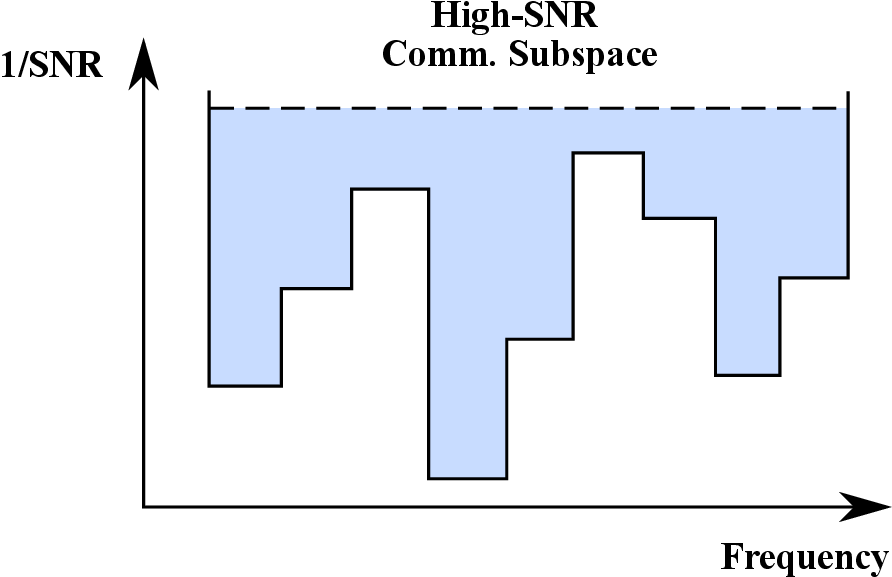}
\end{minipage}
}\\
\subfloat[\scriptsize Sensing Subspace]{
\centering
\includegraphics[width=.65\textwidth]{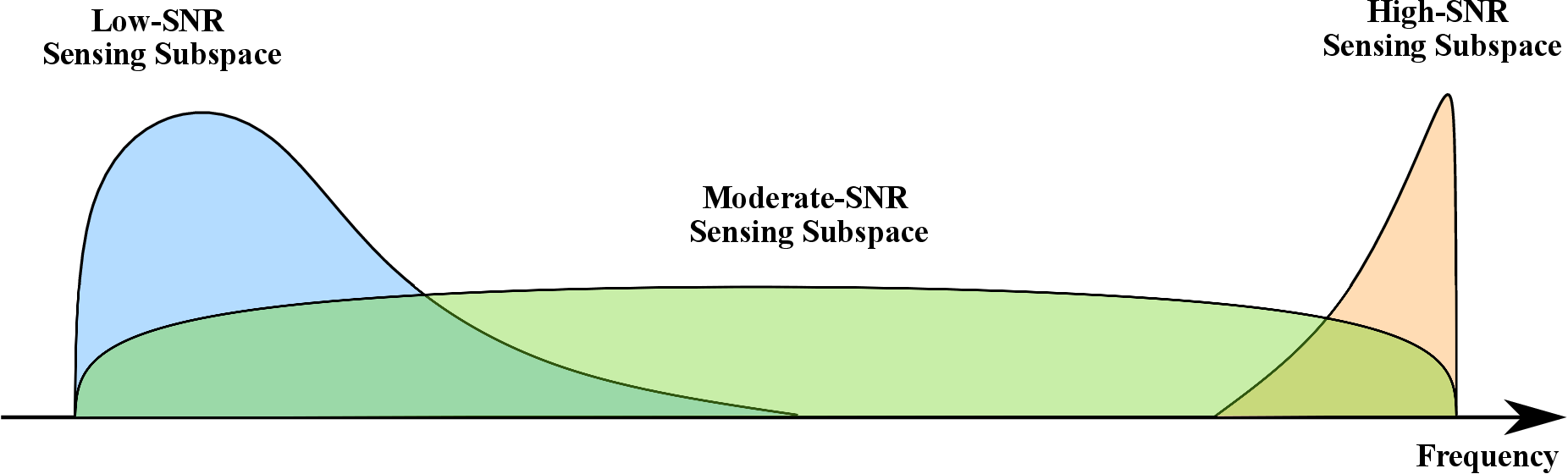}
}
\caption{Frequency-domain communication and sensing subspaces (for the ranging task) in low- and high-\ac{snr} regimes.}
\vspace{-3mm}
\label{fig:st_frequency_domain}
\vspace{-3mm}
\end{figure*}

Note that this is a remarkable observation suggesting that sensing tasks have a unique preference on the subspaces where the \ac{isac} signal resides. This is in stark contrast to communication tasks, for which the optimal frequency-domain power allocation scheme is the water-filling strategy. In particular, the water-filling strategy assigns more power to frequency bands having higher \ac{snr}, and might even abandon some low-\ac{snr} bands when the total power constraint is stringent, or equivalently, when the total \ac{snr} is relatively low. In the language of \ac{st}, we may say that in the frequency domain, the communication subspace corresponds to those frequency bands with high \ac{snr}. By contrast, the sensing subspace is not solely dependent on the \ac{snr}; rather, as the total \ac{snr} increases, the sensing subspace would move from the low-frequency band to the high-frequency band, as portrayed in Fig.~\ref{fig:st_frequency_domain}.

We may obtain further insights from the perspective of resource allocation. One of the resources of communication tasks is the \ac{dof}, which is manifested as the bandwidth in the frequency domain. For communication tasks, the value of the frequency bands only depends on their \emph{quality} (in terms of \ac{snr}). However, for sensing tasks, the value of frequency bands depends not only on their quality, but on their \emph{location} as well. In light of this, we may say that the \ac{dof} is not, in its nature, a sensing resource. We are thus motivated to ask the following question: \emph{What do we really mean when we say ``sensing resources''?}

\section{Concluding Remarks and Open Challenges}
Unfortunately, at the moment of writing, we do not have a  well-stated answer to this question. After all, in contrast to communication tasks always aiming to deliver information, sensing tasks have vastly diverse purposes, and hence may rely on different resources. Apart from this question, many important problems remain open in the context of \ac{snc} tradeoffs. To name a few:
\begin{enumerate}
    \item How does the \ac{drt} manifest itself under generic sensing performance metrics?
    \item How do we characterize the \ac{snc} tradeoff when the channels are not memoryless? What are the performances of online and offline estimators in such scenarios?
    \item How do we design \ac{isac} systems that are capable of achieving the entire capacity-distortion boundary (not just the corner points)?
    \item Can we unify \ac{snc} performance metrics?
\end{enumerate}

These challenging questions remind us that there is still a long way to go before the merit of \ac{isac} can be fully understood and utilized. Nevertheless, the \ac{st}-\ac{drt} decomposition (i.e. the ``projector metaphor'') is likely to be a useful meta-intuition in future investigations of \ac{isac} systems: the fundamental tradeoff in \ac{isac} is manifested as the preference discrepancies between \ac{snc} tasks, concerning both the resources (\ac{st}) and the signal patterns (\ac{drt}).

\bibliographystyle{IEEEtran}
\bibliography{isac,ISAC_overview}

\end{document}